\documentclass{imsart}

\RequirePackage{amsthm,amsmath,amsfonts,amssymb}
\RequirePackage[authoryear]{natbib} 
\RequirePackage[colorlinks,citecolor=blue,urlcolor=blue]{hyperref}
\RequirePackage{graphicx}

\usepackage{amsmath,xcolor}

\usepackage{amssymb}
\usepackage{array}
\usepackage[capitalise]{cleveref}
\usepackage{natbib}
\usepackage{multirow}
\usepackage[justification=centering]{caption}
\usepackage{float}
\usepackage{algorithmic}
\usepackage{algorithm}
\usepackage{booktabs}
\usepackage{makecell}
\usepackage{threeparttable}
\numberwithin{equation}{section}
\makeatletter
\renewcommand*\env@matrix[1][*\c@MaxMatrixCols c]{%
	\hskip -\arraycolsep
	\let\@ifnextchar\new@ifnextchar
	\array{#1}}
\makeatother

\newcommand{\bu}{{\mathbf u}}
\newcommand{\X}{{\mathbf X}}

\newcommand{\A}{{\mathbf A}}

\newcommand{\bP}{{\mathbf P}}
\newcommand{\bS}{{\mathbf S}}

\newcommand{\y}{{\mathbf y}}
\newcommand{\z}{{\mathbf z}}
\newcommand{\bs}{{\mathbf s}}
\newcommand{\I}{{\mathbf I}}

\newcommand{\ba}{{\mathbf a}}

\newcommand{\bt}{{\mathbf t}}
\newcommand{\bT}{{\mathbf T}}
\newcommand{\bx}{{\mathbf x}}

\newcommand{\bv}{{\mathbf v}}

\newcommand{\bq}{{\mathbf q}}

\newcommand{\um}{\underline{m}}

\newcommand{\E}{{\rm E}}
\newcommand{\0}{{\mathbf 0}}
\newcommand{\1}{{\mathbf 1}}
\newcommand{\bxi}{{\boldsymbol \xi}}

\newcommand{\bM}{{\mathbf M}}

\newcommand{\bQ}{{\mathbf Q}}

\newcommand{\tr}{{\text{\rm tr}}}
\newcommand{\supp}{{\text{\rm supp}}}

\newcommand{\balpha}{{\boldsymbol \alpha}}

\newcommand{\bSigma}{{\boldsymbol \Sigma}}

\newcommand{\bsigma}{{\boldsymbol \sigma}}
\newcommand{\bmu}{{\boldsymbol \mu}}
\allowdisplaybreaks[4]

\startlocaldefs
\theoremstyle{plain}

\newtheorem{theorem}{Theorem}[section]
\newtheorem{lemma}{Lemma}[section]
\theoremstyle{definition}

\newtheorem{remark}{Remark}[section]

\endlocaldefs

\begin{document}

\begin{frontmatter}
\title{High-Dimensional Precision Matrix Quadratic Forms: Estimation Framework for $p > n$}
\runtitle{Estimation of Precision Matrix Quadratic Forms}

\begin{aug}
\author[A]{\fnms{Shizhe}~\snm{Hong} \ead[label=e1]{hong.shizhe@163.sufe.edu.cn}},
\author[A]{\fnms{Weiming}~\snm{Li}\ead[label=e2]{li.weiming@shufe.edu.cn}}
\and
\author[B]{\fnms{Guangming}~\snm{Pan}
\ead[label=e3]{gmpan@ntu.edu.sg}}
\address[A]{School of Statistics and Data Science, Shanghai University of Finance and Economics \printead[presep={ ,\ }]{e1,e2}}

\address[B]{School of Physical and Mathematical Sciences, Nanyang Technological University \printead[presep={,\ }]{e3}}
\end{aug}

\begin{abstract}
We propose a novel estimation framework for quadratic functionals of precision matrices in high-dimensional settings, particularly in regimes where the feature dimension $p$ exceeds the sample size $n$. Traditional moment-based estimators with bias correction remain consistent when $p<n$ (i.e., $p/n \to c <1$). However, they break down entirely once $p>n$, highlighting a fundamental distinction between the two regimes due to rank deficiency and high-dimensional complexity. Our approach resolves these issues by combining a spectral-moment representation with constrained optimization, resulting in consistent estimation under mild moment conditions. 

The proposed framework provides a unified approach for inference on a broad class of high-dimensional statistical measures. We illustrate its utility through two representative examples: the optimal Sharpe ratio in portfolio optimization and the multiple correlation coefficient in regression analysis. Simulation studies demonstrate that the proposed estimator effectively overcomes the fundamental $p>n$ barrier where conventional methods fail.
\end{abstract}



\end{frontmatter}

\section{Introduction}

Let \(\bx = (x_1, \ldots, x_p)^\top\) be a random vector in \(\mathbb{R}^p\) with mean \(\bmu\) and covariance matrix \(\bSigma\). 
For a fixed vector \(\ba \in \mathbb{R}^p\), consider the quadratic form
\begin{align}\label{quad}
    \tau_p(\ba) \triangleq \ba^\top \bSigma^{-1} \ba,
\end{align}
which defines the precision-weighted squared norm of \(\ba\) relative to the precision matrix \(\bSigma^{-1}\). 
This fundamental quantity arises in multivariate statistical theory and has many applications across diverse fields through specific instantiations of \(\ba\). In portfolio theory, \(\tau_p(\bmu)\) represents the squared optimal Sharpe ratio while the reciprocal of \(\tau_p(\mathbf{1}_p)\) provides the global minimum variance \citep{Campbell1997}, where $\1_p$ denotes the $p$-dimensional vector of ones; In statistical classification, \(\tau_p(\bx_0 - \bmu)\) measures the squared Mahalanobis distance between \(\bx_0\) and the population mean \citep{DiscriminantAnalysis2004}; In multivariate regression and canonical correlation analysis, \(\tau_p(\ba)\) appears in multiple correlation coefficients where \(\ba\) is a vector of marginal covariances \citep{Anderson03}. Further applications are present in machine learning \citep{MD-SVM2007} and signal detection \citep{SignalProcessing2018}. Despite its ubiquity, $\tau_p$ is rarely directly observable in practice, as the covariance matrix $\bSigma$ typically requires estimation from data, while the vector $\ba$ may be known or unknown depending on application contexts.

For estimating $\tau_p$, the conventional moment method applies when $p < n$, corresponding to the asymptotic regime $p/n \to c \in [0,1)$. Given that $\ba$ is known and $n$ i.i.d.\ observations $\bx_1,\dots,\bx_n$ from the population $\bx$, we construct the estimator as follows. The sample covariance matrix
\begin{align*}
\bS_n = \frac{1}{n-1}\sum_{i=1}^n (\bx_i - \bar{\bx})(\bx_i - \bar{\bx})^\top,
\end{align*}
where $\bar{\bx} = n^{-1}\sum_{i=1}^n\bx_i$, serves as the moment estimator for $\bSigma$. The plug-in estimator $\ba^\top \bS_n^{-1}\ba$ is consistent for $\tau_p$ when $p$ is fixed ($n\to\infty$, $c=0$) but requires bias correction when $p/n \to c \in (0,1)$. Specifically, under this $p < n$ regime, \cite{Bai2007,pan2014comparison} established that
\begin{align*}
\ba^\top \bS_n^{-1}\ba = \frac{\tau_p}{1 - c_n}+o_p(||\ba||^2), \quad c_n \triangleq p/n.
\end{align*}
This enables a consistent estimation via the scaled moment estimator $(1 - c_n)\ba^\top \bS_n^{-1}\ba$. When $\ba$ is unknown and substituted with an estimate, a second-round bias correction becomes necessary while preserving analytical tractability; see \cite{BaiPortfolio2009,Zheng14} for technical details.

\begin{figure}[ht]
\centering
\begin{minipage}[t]{0.5\linewidth}
	\centering
	\includegraphics[width=2.4in]{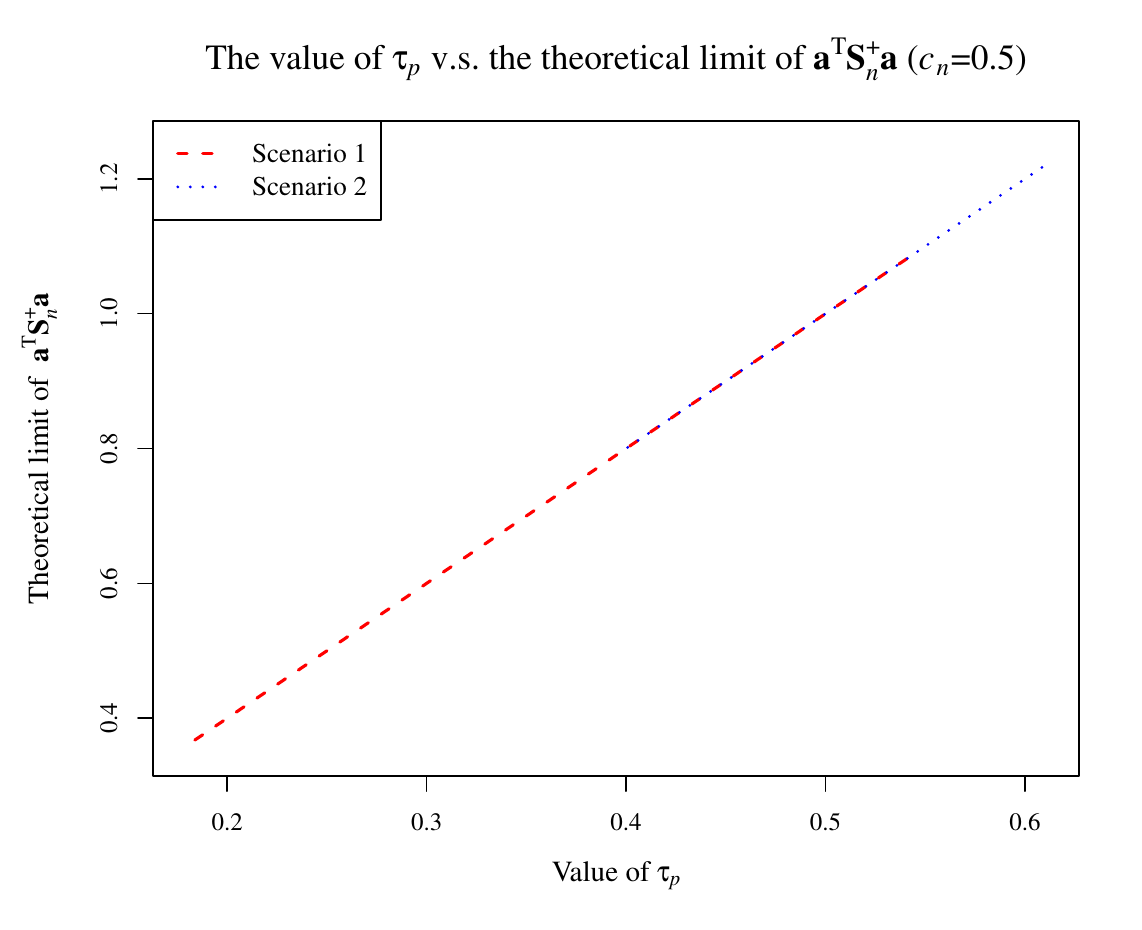}
\end{minipage}%
\begin{minipage}[t]{0.5\linewidth}
	\centering
	\includegraphics[width=2.4in]{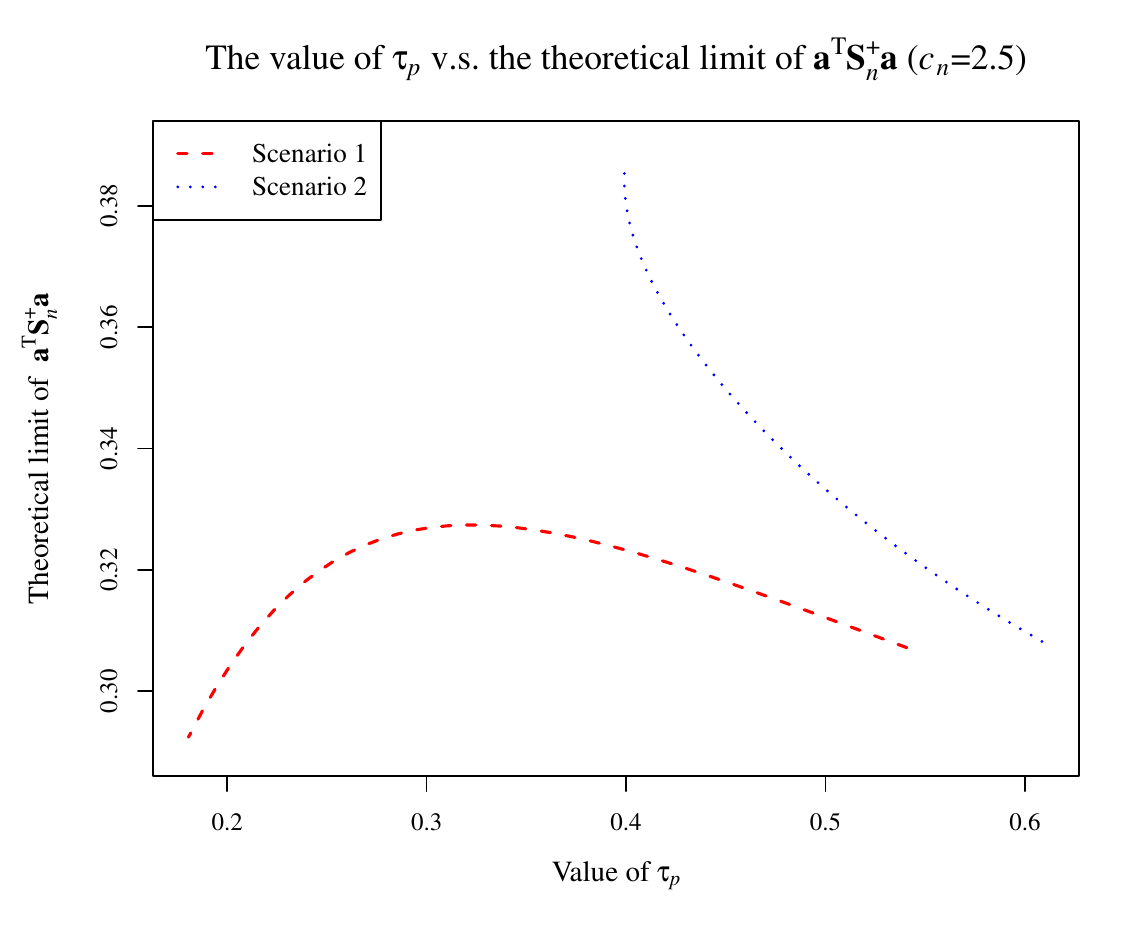}
\end{minipage}
\caption{Comparison between $\tau_p$ and $\ba^\top\bS_n^+\ba$ (theoretical limit) for $p=200$ with 
$\bSigma=\left(r^{|i-j|}\right)_{i,j=1}^p$, where $r$ varies from $0.3$ to $0.7$. 
In Scenario~1, $\ba$ is the uniform vector with all entries equal to $1/\sqrt{p}$; 
in Scenario~2, the first eight entries of $\ba$ are $1/\sqrt{8}$ while the remaining entries are zero.
}
\label{aSna-limit}
\end{figure}

Estimation of $\tau_p$ becomes statistically challenging when $p > n$ due to rank deficiency in the sample covariance matrix $\bS_n$. This limitation becomes evident when substituting the Moore-Penrose inverse $\bS_n^{+}$ into $\tau_p$, which results in
\begin{align}\label{quad-MPinverse}
\ba^\top \bS_n^{+}\ba = m_1\left\{\ba^\top \left(m_0\bSigma+\I\right)^{-1}\ba-\ba^\top \left(m_0\bSigma+\I\right)^{-2}\ba\right\} +o_p(||\ba||^2),
\end{align}
as $c_n\to c\in (1,\infty)$, where $m_0$ and $m_1$ are two positive constants depending on the ratio $c_n$ and the eigenvalues of $\bSigma$ (see Section S.7 in the supplementary material). Crucially, this limit is not a one-to-one function of the target parameter \(\tau_p = \ba^\top \bSigma^{-1} \ba\); consequently, \(\tau_p\) cannot be uniquely recovered from \(\ba^\top \bS_n^{+}\ba\) or its limit. Figure~\ref{aSna-limit} highlights the contrast between the moderate- and high-dimensional regimes. 
When \(p<n\), the relationship between \(\tau_p\) and the limit of \(\ba^\top \bS_n^{-1}\ba\) is injective, so \(\tau_p\) can be recovered in principle. In sharp contrast, when \(p>n\), a fundamental change occurs: the mapping ceases to be injective, and identical limiting values of \(\ba^\top \bS_n^{+}\ba\) may correspond to distinct \(\tau_p\). 
This non-identifiability prevents consistent estimation of \(\tau_p\) using pseudoinverse-based methods and exposes a fundamental barrier to inference in the high-dimensional setting.
Alternative approaches to improving the estimation of $\bSigma^{-1}$ include regularization techniques such as sparsity-based methods with $\ell_0$/$\ell_1$ constraints \citep{friedman2008sparse,CLIME2011,scaleLasso2013,Fan2016,CARE2025}, shrinkage estimators \citep{Ledoit2004,Ledoit2012,Ledoit2017,Ledoit2018}, and approaches developed under factor model structures \citep{Fan2008,Fan2013,Fan2018,Daniele2025}. However, their direct application to quadratic forms, such as $\tau_p$, is problematic: consistency typically relies on sparsity or low-rank assumptions, and plug-in estimators are generally biased, as they fail to recover the inner products between $\ba$ and the eigenvectors of $\bSigma$.

This paper introduces a new framework for estimating the high-dimensional quadratic form $\tau_p$ in the challenging regime where $p>n$. Our approach relies on a spectral reinterpretation of $\tau_p$. 
Specifically, let $\bSigma$ have the spectral decomposition
\begin{align*}
\bSigma = \sum_{i=1}^p \lambda_i \bu_i \bu_i^\top,
\end{align*}
where $\{\lambda_i\}_{i=1}^p$ are the eigenvalues and $\{\bu_i\}_{i=1}^p$ are the corresponding eigenvectors. 
The quadratic form can then be expressed as
\begin{align}\label{inversemoment}
\tau_p = \ba^\top\bSigma^{-1}\ba 
= \sum_{i=1}^p \lambda_i^{-1} (\ba^\top\bu_i)^2 
= \int x^{-1} dF^{\bSigma,\ba}(x),
\end{align}
where the function 
\begin{align*}
F^{\bSigma,\ba}(x) = \sum_{i=1}^p (\ba^\top\bu_i)^2 \mathbb{I}(\lambda_i \leq x)
\end{align*}
is the vector empirical spectral distribution (VESD) of $\bSigma$ relative to $\ba$ \citep{Bai10}. 
This distribution encodes the interaction between $\ba$ and the eigenstructure of $\bSigma$. 
Reformulating $\tau_p$ in terms of $F^{\bSigma,\ba}$ transforms the original estimation problem into one of approximating this spectral measure. 
Unlike the classical moment method, our approach does not require the invertibility of $\bS_n$, thereby overcoming the rank deficiency inherent in the $p>n$ regime. Consequently, a consistent estimator of $F^{\bSigma,\ba}$ yields a consistent plug-in estimator of $\tau_p$.

Our theoretical contribution proceeds in two stages. 
First, using tools from random matrix theory, we develop a comprehensive framework for estimating the VESD in the baseline case where the design vector $\ba$ is assumed known. 
This idealized analysis includes: 
(1) proving the probabilistic convergence of sample VESD statistics constructed from $\bS_n$ and $\ba$ under mild moment conditions; 
(2) obtaining consistent estimators for all moments of $F^{\bSigma,\ba}$ via complex analysis; and 
(3) reconstructing the population VESD by combining these moment estimates with constrained linear programming. 
Second, we extend the framework to the more realistic setting where $\ba$ is unknown by introducing a systematic bias-correction scheme. 
This extension yields valid statistical procedures for portfolio optimization and high-dimensional correlation analysis when $\ba$ is replaced by its plug-in estimator.

The paper is organized as follows.
Section \ref{results} develops the estimation framework and its theoretical properties under the assumption that $\ba$ is known.
Section \ref{applications} extends the framework to the more realistic case where $\ba$ is unknown, and illustrates it through two applications: (i) estimation of the optimal Sharpe ratio, and (ii) inference on multiple correlation in regression.
All proofs and additional technical results are deferred to the supplementary material.

\section{Quadratic form estimation with known $\ba$}\label{results}
\subsection{Model and assumptions}
This section introduces the theoretical framework for estimating the quadratic form $\tau_p = \ba^\top \bSigma^{-1} \ba$ in the baseline case where the vector $\ba$ is known. 
At the population level, we adopt the location–scale decomposition
\begin{align}\label{IC}
\bx = \bmu + \A\z,
\end{align}
which holds for any random vector with finite second moments, where
\begin{itemize}
\item[(1)] $\bmu \in \mathbb R^p$ is the mean vector of $\bx$, and $\A \in \mathbb{R}^{p \times p}$ is a full-rank matrix such that $\A\A^\top = \bSigma$ (the population covariance matrix);
\item[(2)] $\z = (z_1, \dots, z_p)^\top \in \mathbb{R}^p$ is the standardized version of $\bx$ with $\E(\z) = \0$ and $\E(\z\z^\top) = \I_p$.
\end{itemize}
Our main assumptions are stated below.

\medskip
\noindent\textbf{Assumption (a)}.  
The dimensions $(p,n)$ tend to infinity
in such a way that 
\begin{align}\label{high-dimensional-regime}
p=p_n=O(n)\quad\text{and}\quad c_n= p/n\to c \in(0,\infty).
\end{align}

\medskip
\noindent\textbf{Assumption (b)}.  For any deterministic matrix $\bQ\in \mathbb R^{p\times p}$ with bounded spectral norm,
\begin{align*}
\E\left|\z^\top \bQ \z - \tr\bQ \right|^2 = o(p^2).
\end{align*}

\medskip
\noindent\textbf{Assumption (c)}.  
The eigenvalues of $\bSigma$ are uniformly bounded away from zero and infinity, i.e., there exist constants $a$ and $b$ such that
\begin{align*}
0<a\leq\liminf_{p\to\infty}\lambda_{\min}\left(\bSigma\right)\leq\limsup_{p\to\infty}\lambda_{\max}\left(\bSigma\right)\leq b<\infty.
\end{align*}

\medskip
\noindent\textbf{Assumption (d)}. For any deterministic unit vector $\bt \in \mathbb{R}^p$ ($\|\bt\| = 1$), 
\begin{align}\label{quad-assumption}
\E\left|\z^\top\bt\right|^4 = O(1).
\end{align}

For simplicity, we normalize $\|\ba\|=1$. 
This does not entail a loss of generality: writing $\ba=\|\ba\|\cdot \ba_0$ with $\|\ba_0\|=1$ gives $\tau_p(\ba)=\|\ba\|^2 \tau_p(\ba_0)$, so results for general $\ba$ follow immediately by rescaling whenever $\|\ba\|$ is bounded. 

\begin{remark}
Assumption (a) specifies the high-dimensional asymptotic regime, which covers the case $p > n$ as a particular instance. Assumption (b) imposes only mild structural conditions, allowing for general dependence among the components of $\z$, and is consistent with the framework for the analysis of eigenvalue distributions in \cite{BaiZhou08}. Assumption (c) guarantees that the spectrum of $\bSigma$ remains well-conditioned, avoiding both degeneracy and divergence as $p$ grows. Assumption (d) bounds the fourth moments of linear projections of $\z$, ensuring the concentration properties needed for quadratic form analysis. 
\end{remark}

\subsection{Convergence of sample VESD}

We begin by recalling two fundamental concepts in random matrix theory. For any Hermitian matrix $\bT \in \mathbb{R}^{p \times p}$ with spectral decomposition $\bT = \sum_{i=1}^p \lambda_i^{\bT} \bxi_i \bxi_i^\top$:
\begin{itemize}
\item[1.] The empirical spectral distribution (ESD) of $\bT$ is defined as
\begin{align*}
F^{\bT}(x) = \frac{1}{p} \sum_{i=1}^p \mathbb{I}(\lambda_i^{\bT} \leq x),
\end{align*}
where $\mathbb{I}(\cdot)$ is the indicator function.

\item[2.] For any unit vector $\bt \in \mathbb{R}^p$ ($\|\bt\|=1$), the vector ESD (VESD) of $\bT$ with respect to $\bt$ is given by
\begin{align*}
F^{\bT,\bt}(x) = \sum_{i=1}^p w_i^{\bt} \mathbb{I}(\lambda_i^{\bT} \leq x), 
\end{align*}
where $w_i^\bt \triangleq (\bt^\top \bxi_i)^2$ are the projection weights.
\end{itemize}

The convergence properties of the ESD $F^{\bS_n}$ for sample covariance matrices $\bS_n$ have been extensively studied since the seminal work of \cite{MP67}. Important extensions include \cite{S95}, which established the convergence under finite second moment conditions with linear dependence structures, and \cite{BaiZhou08}, which proved the convergence under general dependence structures, as specified in Assumption (b). Our analysis will primarily use the results from \cite{BaiZhou08}, where the convergence is characterized by the Stieltjes transform of $F^{\bS_n}$,
\begin{align*}
m_n(z)\triangleq \int\frac{1}{x-z}dF^{\bS_n}(x), \quad z\in\mathbb C^+\triangleq\left\{z\in\mathbb C:\Im(z)>0\right\}.
\end{align*}
The set $\mathbb C^+$ denotes the upper complex plane. 

\begin{lemma}[\cite{BaiZhou08}]\label{lem1}
Under Assumptions (a)-(b)-(c), the Stieltjes transform $m_n(z)$ of $F^{\bS_n}$ converges almost surely, that is,
\begin{align}\label{convergence}
m_n(z)-m(z)\xrightarrow{a.s.}0,\quad \forall z\in\mathbb C^+,
\end{align}
where $m(z)$ is the unique solution to the equation
\begin{align}\label{MPeq0}
m(z)=\int \frac{1}{x(1-c_n-c_n z m(z))-z} d F^{\bSigma}(x)
\end{align}    
in the set $\{z\in \mathbb C^+: -(1-c_n)/z+c_nm(z)\in\mathbb{C}^+\}$.
\end{lemma}
\begin{remark}
Lemma \ref{lem1} presents the almost sure convergence of the Stieltjes transform $m_n(z)$. Through the inversion theorem for Stieltjes transforms, this lemma guarantees the weak convergence of the ESD $F^{\bS_n}$ to a limiting distribution. In addition,
the companion Stieltjes transform
\begin{align*}
\um(z) \triangleq -\frac{1-c_n}{z} + c_n m(z)
\end{align*}
converts the fixed-point equation \eqref{MPeq0} into another canonical form
\begin{align}\label{MPeq}
z = -\frac{1}{\um(z)} + \int \frac{c_n x}{1 + \um(z)x} dF^\bSigma(x),
\end{align}
from which the support of the limiting spectral distribution can be derived \citep{SC95}. We refer to this equation as the Mar\v{c}enko–Pastur (MP) equation.
\end{remark}

Next, we investigate the convergence of the VESD $F^{\bS_n,\ba}$ for the sample covariance matrix $\bS_n$ with respect to a deterministic vector $\ba$. Its Stieltjes transform is given by
\begin{align}\label{snz}
s_n(z) \triangleq \int \frac{1}{x-z} dF^{\bS_n,\ba}(x), \quad z \in \mathbb{C}^+.
\end{align}

\begin{theorem}\label{th-convergence-s}
Under Assumptions (a)-(b)-(c)-(d), the Stieltjes transform $s_n(z)$ of $F^{\bS_n,\ba}$ converges in probability, that is,
\begin{align}\label{convergence-s}
s_n(z) - s(z) \xrightarrow{i.p.} 0 \quad \forall z \in \mathbb{C}^+,
\end{align}
where 
\begin{align}\label{sz}
s(z) &= \ba^\top \left(-z\I_p - z\um(z)\bSigma\right)^{-1} \ba \
= \int \frac{1}{-z - z\um(z)x} dF^{\bSigma,\ba}(x),
\end{align}
and $\um(z)$ is the companion Stieltjes transform defined in the MP equation \eqref{MPeq}.
\end{theorem}

\begin{remark}
Theorem \ref{th-convergence-s} shows that the Stieltjes transform $s_n(z)$ converges, which by the inversion theorem implies the weak convergence of the VESD $F^{\bS_n,\ba}$. This extends the seminal work of \cite{Bai2007} from the i.i.d.\ setting to more general dependence structures in $\z$. The conclusion also holds when the sample covariance $\bS_n$ in the VESD is replaced by robust scatter estimators, such as Tyler's $M$-estimator \citep{Tyler1987} or the spatial-sign covariance matrix \citep{locantore1999robust}, under elliptical distributions. Further discussion can be found in \cite{BaiZhou08}.
\end{remark}

\subsection{Estimation of the moments of $F^{\bSigma,\ba}$}

We now study the estimation of the moments of the VESD $F^{\bSigma,\ba}$. The $j$-th moment is
\begin{align*}
\alpha_j \triangleq \int x^j   dF^{\bSigma,\ba}(x) = \ba^\top\bSigma^j\ba, \quad j \in \mathbb N,
\end{align*}
where $\mathbb N$ denotes the set of positive integers. Accurate estimation of these moments is fundamental to the VESD analysis.

Theorem \ref{th-convergence-s} reveals that while the sample VESD $F^{\bS_n,\ba}$ deviates from $F^{\bSigma,\ba}$ in high dimensions, their connection is preserved through the link function \eqref{sz}. By complex analytic techniques, we obtain an exact moment-reconstruction formula
\begin{align}\label{moment-alphaj}
\alpha_j
	=&(-1)^j\frac{1}{2\pi \mathrm i}\oint_\mathcal{C}\frac{zs(z)\um'(z)}{\um^{j+1}(z)}dz, \quad j \in \mathbb N,
\end{align}
where $\um'(z)$ denotes the derivative of the function $\um(z)$ and the contour 
$\mathcal{C}$ is positively oriented, enclosing the support of the limiting spectral distribution of $F^{\bS_n}$.
This representation expresses population moments directly through the Stieltjes transforms $\um(z)$ and $s(z)$. Substituting their sample counterparts yields the estimator
\begin{align}\label{hatalpha}
\hat\alpha_j
	= (-1)^j\frac{1}{2\pi \mathrm i}\oint_\mathcal{C}\frac{z s_n(z)\um_n'(z)}{\um_n^{ j+1}(z)} dz, \quad j \in \mathbb N,
\end{align}
where
\begin{align*}
\um_n(z) \triangleq -\frac{1-c_n}{z} + c_n m_n(z).
\end{align*}

To analyze the asymptotic properties of $\hat{\alpha}_j$, we require spectral norm control on $\bS_n$ to ensure valid contour integration.

\medskip
\noindent\textbf{Assumption (b$^*$)}. The random vector $\z$ satisfies either:
\begin{itemize}
\item[(i)] Light-tailed independence: $\{z_j\}_{j=1}^p$ are i.i.d.\ with $\E|z_1|^4 < \infty$; or
\item[(ii)] Log-concavity: $\z$ has a log-concave density.
\end{itemize}

\begin{remark}
These conditions provide sufficient spectral control in high-dimensional settings. They guarantee $||\bS_n|| = O(1)$ almost surely as $p,n \to \infty$
and thus ensure the existence of a fixed contour $\mathcal{C}$ that encloses all eigenvalues of $\bS_n$ for large $p, n$. Also note that under this assumption, both Assumptions (b) and (d) are automatically satisfied.
\end{remark}

\begin{theorem}\label{moments}
Under Assumptions (a), (b$^*$), (c), the moment estimator $\hat{\alpha}_j$ satisfies
\begin{equation}
\hat{\alpha}_j - \alpha_j \xrightarrow{i.p.} 0, \quad \text{as } n,p \to \infty,
\end{equation}
for any fixed integer $j \in \mathbb{N}$. 
\end{theorem}

\begin{remark}
Theorem \ref{moments} establishes the consistency of the moment estimator $\hat{\alpha}_j$. 
In practice, $\hat{\alpha}_j$ can be computed via contour integration, where Cauchy's residue theorem yields explicit formulas by evaluating residues at the poles determined by the zeros of $\um_n(z)$ and the eigenvalues of $\bS_n$. 
Closed-form expressions exist for moments of all orders, but their complexity grows rapidly with $j$. 
For example, the first moment admits a relatively simple formula:
\begin{align}
\hat{\alpha}_1 = n \ba^\top \bS_n \ba - \sum_{i=1}^\psi \frac{\eta_i s_n'(\eta_i) + s_n(\eta_i)}{\um_n'(\eta_i)},
\end{align}
where $\psi = \min\{p,n-1\}$ and $\eta_1>\cdots>\eta_\psi$ are the zeros of $\um_n(z)$. 
For higher-order moments ($j \geq 2$), the formulas involve lengthy sums of derivatives up to order $j$.  
To handle this complexity, we provide Mathematica code in Appendix \ref{sec-hatalpha}, which can generate the exact symbolic expressions.
\end{remark}

\subsection{VESD Estimation}\label{VESD-est}

To estimate the VESD $F^{\bSigma,\ba}$, we employ a moment-matching method introduced by \cite{Kong17}, originally designed for the ESD estimation. The basic idea is to approximate the target distribution by a discrete measure on a fine grid, with weights determined by matching the first $k$ moments, where $k$ is a tuning parameter controlling the number of moments used. Since the approach relies solely on moment estimates, it extends naturally to the VESD setting. For completeness, the procedure is summarized below.

\begin{algorithm}[H]
\caption{Estimation of the VESD $F^{\bSigma,\ba}$}
\label{alg}
\begin{algorithmic}
\STATE {\bf Step 1.} Select a tuning parameter $k$. 
Compute the estimates of the first $k$ moments $\hat\balpha = (\hat\alpha_1,\dots,\hat\alpha_k)^\top$ using \eqref{hatalpha}.

\STATE {\bf Step 2.} Choose two numbers $0 < a_0 < b_0 < \infty$ such that the interval $(a_0,b_0)$ contains the closure of $\cup_{p=1}^\infty \supp(F^{\bSigma,\ba})$\footnotemark, where $\supp(F^{\bSigma,\ba})$ denotes the support of $F^{\bSigma,\ba}$. 
Define the grid points on $[a_0,b_0]$ with step size $h \leq 1/\max\{n,p\}$:
\begin{align*}
d_i = a_0 + (i-1)h,\quad i=1,\dots,t \quad\text{with}\quad t = \Big\lfloor \frac{b_0-a_0}{h} \Big\rfloor + 1.
\end{align*}

\STATE {\bf Step 3.} Solve the following linear program for $\bq = (q_1,\ldots,q_t)^\top \in \mathbb{R}^t$:
\[
\hat\bq = \arg\min_{\bq} \|\bM \bq - \hat\balpha\|_1 
\quad \text{subject to} \quad \bq \ge 0,\ \mathbf{1}_t^\top\bq = 1,
\]
where the $(i,j)$-th entry of $\bM$ is given by $d_j^i$.

\STATE {\bf Step 4.} Construct the estimator:
\[
\hat{F}^{\bSigma,\ba}(x) = \sum_{i=1}^t \hat q_i  \mathbb{I}\{d_i \le x\},
\quad \text{where } \hat\bq = (\hat q_1,\ldots,\hat q_t)^\top.
\]
\end{algorithmic}
\end{algorithm}

\footnotetext{ Note that $F^{\bSigma,\ba}$ is defined for each dimension $p$.}

\begin{theorem}\label{th-W1distance}
Suppose that Assumptions (a)-(b*)-(c) hold. 
If, in addition, the tuning parameter $k=k_n$ satisfies 
\begin{align}\label{kinfty}
k_n \to \infty, \quad \frac{k_n}{\log n} \to 0,
\end{align}
then we have
\begin{align*}
W_1(\hat{F}^{\bSigma,\ba},F^{\bSigma,\ba}) \xrightarrow{i.p.} 0,
\end{align*}
where $W_1(\cdot,\cdot)$ denotes the 1-Wasserstein distance between distribution functions.
\end{theorem}

\begin{remark}
The growth condition \eqref{kinfty} balances two sources of error: the approximation error of representing $F^{\bSigma,\ba}$ with finitely many moments, and the estimation error of high-order moments from the data. Increasing $k_n$ reduces the approximation error, while restricting its growth rate controls the variance inflation inherent in estimating higher-order moments.
\end{remark}

\subsection{Inference for high-dimensional quadratic forms}\label{sec:quad}

Quadratic forms of the type
\begin{align*}
\tau_p = \int \frac{1}{x} \, dF^{\bSigma,\ba}(x) = \ba^\top \bSigma^{-1}\ba
\end{align*}
play a fundamental role in high-dimensional statistics.  
Given a VESD estimator $\hat F^{\bSigma,\ba}$ obtained from Algorithm~\ref{alg}, a natural estimator of $\tau_p$ is the corresponding plug-in functional
\begin{align}\label{eq:tauhat}
\hat\tau_p = \int \frac{1}{x} d\hat F^{\bSigma,\ba}(x).
\end{align}
From Assumption (c),
$\tau_p$ is a continuous functional of $F^{\bSigma,\ba}$ under the Wasserstein distance $W_1$.  
By Theorem~\ref{th-W1distance}, the estimator $\hat F^{\bSigma,\ba}$ is consistent and therefore $\hat\tau_p$ is also consistent.

\vspace{6pt}
\noindent\textbf{Finite-sample instability.}  
Despite its asymptotic validity, the naive plug-in estimator, $\hat\tau_p$, may be unstable in finite-sample situations.  Algorithm~\ref{alg} reconstructs $\hat F^{\bSigma,\ba}$ by solving a linear program (LP) that matches finitely many empirical moments.  
In practice, high-order moment estimates can fluctuate severely, occasionally producing inadmissible values (e.g., negative estimates).  
These instabilities propagate through the LP, resulting in a rough $\hat F^{\bSigma,\ba}$ and a noisy $\hat\tau_p$.

\vspace{6pt}
\noindent\textbf{Stabilizing moment inputs.}  
To enhance stability without altering asymptotic properties, we regularize the moment estimates before feeding them into the LP.  
Using simple Jensen-type inequalities,
\begin{align}\label{jensen-ineq}
\alpha_1^j\leq \alpha_j\quad\text{and}\quad a_0^j < \alpha_j < b_0^j, \quad j\ge1,
\end{align}
we construct truncated moment estimators
\begin{align*}
\hat{\alpha}_{1,\mathrm{tr}} &\triangleq \min\bigl\{\max\{\hat\alpha_1, a_0\}, b_0\bigr\},\\
\hat{\alpha}_{j,\mathrm{tr}} &\triangleq \min\bigl\{\max\{\hat\alpha_j, \hat{\alpha}_{1,\mathrm{tr}}^j-\delta\}, b_0^{ j}\bigr\},\quad j\ge2,
\end{align*}
where $\delta>0$ is an arbitrarily small constant introduced to ensure strict inequalities.
Additionally, the $j$-th moment constraint in the LP is weighted by $1/\hat\alpha_{j,\mathrm{tr}}$ to reduce the influence of variability in higher-order moments. Using these stabilized inputs yields a more reliable VESD estimator and, consequently, a more stable plug-in estimator of $\tau_p$, referred to as
\[
\hat\tau_{p,\mathrm{stab}}
    \triangleq
    \int \frac{1}{x}  d\hat F^{\bSigma,\ba}(x)
     \Big|_{\hat\bq \leftarrow \hat\bq_{\mathrm{stab}}},
\]
where $\hat\bq_{\mathrm{stab}}$ is obtained by solving the LP:
\[
\hat\bq_{\mathrm{stab}}
    = \arg\min_{\bq}
      \left\|
        (\bM\bq) \oslash \hat\balpha_{\mathrm{tr}} - \mathbf{1}_k
      \right\|_1
    \quad\text{subject to}\quad
    \bq \ge 0,\ \mathbf{1}_t^\top\bq = 1,
\]
with  $\bM$ defined in Algorithm~\ref{alg}, $\hat\balpha_{\mathrm{tr}} = (\hat{\alpha}_{1,\mathrm{tr}},\ldots,\hat{\alpha}_{k,\mathrm{tr}})^\top$, and
$``\oslash"$ denoting element-wise division.

\begin{theorem}\label{th:tau}
Under Assumptions (a), (b$^*$), and (c), and if $k=k_n$ satisfies $k_n\to\infty$ and $k_n/\log n\to0$, both the naive and stabilized estimators are consistent, i.e.,
\[
\hat\tau_p - \tau_p \xrightarrow{i.p.} 0 
\quad\text{and}\quad 
\hat\tau_{p,\mathrm{stab}} - \tau_p \xrightarrow{i.p.} 0,
\]
as $n,p\to\infty$.
\end{theorem}

\begin{remark}
Theorem~\ref{th:tau} shows that $\tau_p$ can be consistently estimated without matrix inversion, using only the moment-based VESD estimator.  
The same framework applies to other smooth linear functionals of $F^{\bSigma,\ba}$.  
In addition, while the naive plug-in estimator is consistent, employing truncated and weighted moments preserves asymptotic validity and can enhance finite-sample stability.
\end{remark}

We conduct simulations to assess the finite-sample performance of the proposed estimators 
$\hat\tau_p$ and $\hat\tau_{p,\mathrm{stab}}$.
Data are generated as $\bx = \bSigma^{1/2} \z$, where $\z$ consists of i.i.d.\ standard normal entries. 
We consider two covariance structures:  
\begin{itemize}  
\item[]{\bf Case 1.} Diagonal matrix $\bSigma = \mathrm{diag}(\sigma_{11},\ldots,\sigma_{pp})$, with $\sigma_{ii}=2.5+2i/p,i=1,\dots,p$;  
\item[]{\bf Case 2.} Band matrix $\bSigma=(\sigma_{ij})$ defined by
\[
\sigma_{ii}=2.5,\quad i=1,\dots,p;\quad 
\sigma_{i,i\pm 1}=0.8,\quad i=1,\dots,p-1,
\]
and $\sigma_{ij}=0$ otherwise.
\end{itemize}
For the design vector $\ba$, we examine two representative settings:  
\begin{itemize}
\item[]{\bf Dense setting 1:} The first $p/2$ entries are $\sqrt{0.8}/\sqrt{p}$ and the remaining $p/2$ are $\sqrt{1.2}/\sqrt{p}$.
\item[]{\bf Sparse setting 1:} The first eight entries are $1/\sqrt{8}$ and the rest are zero.  
\end{itemize} 
These combinations of covariance structures and design vectors allow us to assess the performance of the estimators under both heterogeneity and dependence in $\bSigma$, as well as varying degrees of sparsity in $\ba$.

In the linear program, we fix the support interval at $(a_0,b_0)=(0.3,5)$, use $k=4$ moments and $h=1/p$ step size, and set the truncation threshold to $\delta=0.01$ for computing $\hat\tau_{p,\mathrm{stab}}$.  
The dimensional ratio is chosen as $c_n \in \{1.25,1.5\}$. All results are based on 5000 repetitions.

Table \ref{t1} reports the number of occurrences of negative moment estimates obtained using \eqref{hatalpha} for sample sizes $n=200,400$ and $800$. 
The results show that moment truncation is essential for constructing $\hat\tau_{p,\mathrm{stab}}$ when both $p$ and $n$ are small, with the issue most evident in Case 1.  
Table \ref{t2} presents the empirical biases and variances of $\hat\tau_{p,\mathrm{stab}}$ for $n=400, 800,$ and $1600$. As $n$ and $p$ increase, both bias and variance decrease toward zero, providing empirical support for the consistency of $\hat\tau_{p,\mathrm{stab}}$.

\begin{table}[h!]
\centering
\captionsetup{font={small}}
\caption{Number of negative moment estimates obtained using \eqref{hatalpha} from 5000 replications.}
\label{t1}
\begin{tabular}{cccccccc} \Xhline{2pt}
\multirow{2}{*}{$c_n$}    & \multirow{2}{*}{Case} & \multicolumn{3}{c}{Dense setting 1} & \multicolumn{3}{c}{Sparse setting 1} \\ \cmidrule(r){3-5} \cmidrule(r){6-8}
&  & $n=200$  & $n=400$  & $n=800$  & $n=200$   & $n=400$  & $n=800$  \\ \hline
\multirow{2}{*}{1.25} & Case 1 & 479     & 118     & 8       & 1657      & 1094      & 702       \\
& Case 2 & 45      & 0      & 0       & 49      & 1      & 0       \\ \hline
\multirow{2}{*}{1.5}  & Case 1 & 605    & 204     & 26      & 1762      & 1335      & 911       \\
& Case 2 & 56      & 2      & 0       & 101      & 6      & 0  \\ \Xhline{2pt}   
\end{tabular}
\end{table}

\begin{table}[htbp]
\centering
\captionsetup{font={small}}
\caption{Empirical biases (variances) of $\hat\tau_{p,\mathrm{stab}}$ from 5000 replications.}
\label{t2}
\begin{tabular}{ccccc} \Xhline{2pt}
\multirow{2}{*}{$(c_n,n)$} & \multicolumn{2}{c}{Case 1}      & \multicolumn{2}{c}{Case 2}      \\ \cline{2-5}
& Dense setting 1  & Sparse setting 1 & Dense setting 1 & Sparse setting 1 \\ \hline
(1.25,400)              & 0.0348(0.0068) & 0.0367(0.0045) & 0.0214(0.0026) & 0.0194(0.0031) \\
(1.25,800)              & 0.0220(0.0025) & 0.0258(0.0028) & 0.0128(0.0008) & 0.0096(0.0012) \\
(1.25,1600)             & 0.0148(0.0012) & 0.0164(0.0011) & 0.0082(0.0003) & 0.0058(0.0007) \\
(1.5,400)               & 0.0378(0.0083) & 0.0408(0.0066) & 0.0244(0.0033) & 0.0226(0.0049) \\
(1.5,800)               & 0.0249(0.0033) & 0.0286(0.0029) & 0.0144(0.0011) & 0.0121(0.0015) \\
(1.5,1600)              & 0.0161(0.0015) & 0.0192(0.0015) & 0.0091(0.0004) & 0.0065(0.0007) \\ \Xhline{2pt}
\end{tabular}
\end{table}

\section{Estimation of $\tau_p$ when $\ba$ is unknown}\label{applications}
\subsection{Estimation framework}\label{sec:3-1}

When the vector $\ba$ is unknown, it is often convenient to reparameterize $\tau_p$ as
\[
    \tau_p = \kappa_{\ba}\cdot \ba_0^\top \bSigma^{-1} \ba_0,
    \quad \text{where} \quad
    \kappa_{\ba} = \|\ba\|^2,
    \quad 
    \ba_0 = \frac{\ba}{\|\ba\|}.
\]
This reparameterization separates the scale $\kappa_{\ba}$ from the direction $\ba_0$, 
which not only improves numerical stability in high-dimensional settings but also facilitates consistent estimation.
The scalar $\kappa_{\ba}$ can then be estimated separately, while $\ba_0$ is approximated by its plug-in estimator 
$\hat\ba_0 = \hat\kappa_{\ba}^{-1/2}\,\hat\ba$.

With this setup, replacing the unknown direction vector $\ba_0$ with its estimator $\hat\ba_0$ in the statistic $\hat\tau_p$
generally introduces a non-negligible bias, which arises from the estimation error in the moment estimators
$\{\hat\alpha_j\}$ defined in \eqref{hatalpha}. 
These estimators involve the Stieltjes transform $s_n(z)$, which depends on the true value of $\ba_0$ and is not observable. 
Since $s_n(z)$ is unavailable, we must base our analysis on the observable quantity
\begin{align*}
\hat s_n(z) \triangleq \hat\ba_0^\top\!\left(\bS_n - z\I_p\right)^{-1}\!\hat\ba_0.
\end{align*}
However, $\hat s_n(z)$ is generally not consistent; that is, $\hat s_n(z) - s_n(z) \not\to 0$ for $z \in \mathbb{C}^+$. 
To address this issue, we analyze the limiting behavior of $\hat s_n(z)$ and derive its explicit relationship with $s_n(z)$. 
Once this relationship is established, we correct the resulting bias in a principled way, and the estimation procedure for $\tau_p$ developed in Section~\ref{results} can then be applied to construct the final estimator.

Although the overall bias-correction framework is unified, its implementation is problem-specific and requires separate derivations.  
We illustrate this framework in the following two representative applications,
estimating the optimal Sharpe ratio and the multiple correlation coefficient, where the relationship between $\hat s_n(z)$ and $s_n(z)$ turns out to be linear, making the bias correction especially simple.

\subsection{Estimating the optimal Sharpe ratio}\label{SRsec}
The Sharpe ratio, rooted in the mean–variance paradigm \citep{portfolio1952}, is a fundamental measure of risk-adjusted return in portfolio theory that captures the trade-off between expected excess return and volatility. 
Under a fixed-risk constraint and assuming the asset return vector follows the model \eqref{IC}, the squared optimal Sharpe ratio (also called the clairvoyant Sharpe ratio) is
\[
\theta_p = \bmu^\top \bSigma^{-1} \bmu.
\]
Accurate estimation of $\theta_p$ is crucial for both risk-adjusted performance evaluation and portfolio optimization. 
Various methods have been proposed for estimating $\theta_p$ in high-dimensional settings, ranging from approaches developed for the classical regime $p<n$ to those applicable when $p>n$ under additional structural assumptions \citep{KanZhou2007,BaiPortfolio2009,Karoui2010,Ao2018,fan2021optimal}. For recent developments, see \cite{Lu2024,Kan2024,Meng2025}, among others.

In this section, we apply the proposed framework to estimate the optimal Sharpe ratio $\theta_p$ without imposing restrictive structural assumptions.  
Since $\theta_p$ coincides with the quadratic form $\tau_p$ when $\ba = \bmu$, the results in Section~\ref{results} provide the theoretical foundation, and we incorporate the bias-correction procedure from Section~\ref{sec:3-1} to account for the estimation of $\ba$ by the sample mean $\hat\ba = \bar\bx$.

Since $\theta_p$ involves an unrestricted Euclidean norm of $\bmu$ (assuming $\bmu \neq \mathbf{0}$), 
we adopt the same reparameterization as in Section~\ref{sec:3-1},
\[
\theta_p = \kappa_{\bmu} \cdot \bmu_0^\top \bSigma^{-1} \bmu_0,
\quad \text{where} \quad
\kappa_{\bmu} = \|\bmu\|^2,\quad
\bmu_0 = \frac{\bmu}{\|\bmu\|}.
\]
The two components are then estimated by
\[
\hat\kappa_{\bmu} = \left|\frac{1}{n(n-1)}\sum_{i \neq j} \bx_i^\top \bx_j\right|,
\quad
\hat\bmu_0 = \hat\kappa_{\bmu}^{-1/2} \, \bar\bx.
\]
To estimate $\bmu_0^\top \bSigma^{-1} \bmu_0$, we first derive and adjust for the bias in $\hat s_n(z)$.  
An explicit calculation shows that
\begin{align*}
\hat s_n(z) 
&= \hat\bmu_0^\top \left(\bS_n - z\I_p\right)^{-1} \hat\bmu_0  \\
&= \bmu_0^\top \left(\bS_n - z\I_p\right)^{-1} \bmu_0 - \hat\kappa_{\bmu}^{-1} \cdot \frac{1 + z \um_n(z)}{z \um_n(z)} + o_p(1).
\end{align*}
Therefore, we define the following bias-adjusted function
\[
s_{n,\mathrm{SR}}(z)\triangleq \hat\kappa_{\bmu}^{-1}\left[\bar\bx^\top\left(\bS_n-z\I_p\right)^{-1}\bar\bx+\frac{1+z\um_n(z)}{z\um_n(z)}\right],
\]
which is then used to construct the moment estimators (see \eqref{moment-alphaj} on Page 6)
\begin{align}\label{intSR}
\hat\alpha_{j,\mathrm{SR}}
= (-1)^j \frac{1}{2\pi\mathrm i}
   \oint_{\mathcal C} 
   \frac{z s_{n,\mathrm{SR}}(z) \um_n'(z)}{\um_n^{j+1}(z)} dz,\quad j\geq 1.
\end{align}
Together with Algorithm~\ref{alg}, this yields an estimator $\hat{F}^{\bSigma,\bmu_0}$ of the VESD $F^{\bSigma,\bmu_0}$.  
Finally, the estimator of $\theta_p$ is given by
\[
\hat\theta_p
  = \hat\kappa_{\bmu}
    \int x^{-1}   d\hat{F}^{\bSigma,\bmu_0}(x).
\]

\begin{theorem}\label{th-thetap}
Assume the conditions of Theorem~\ref{th-W1distance} hold and that $\|\bmu\|$ is bounded. Then, we have
\[
\hat\theta_p - \theta_p \xrightarrow{i.p.} 0,
\]
as $n, p \to \infty$.
\end{theorem}

\begin{remark}
Theorem \ref{th-thetap} establishes the consistency of the squared Sharpe ratio estimator $\hat\theta_p$.
For practical implementation, Appendix~\ref{residueSR} provides Mathematica code for generating the residue-based moment estimators in \eqref{intSR}.
\end{remark}

\begin{remark}[Extension to the mean-variance frontier]
The squared optimal Sharpe ratio $\theta_p$ is one of the key quantities characterizing the mean-variance frontier (MVF) \citep{Merton1972}.  
The MVF is fully characterized by the following key quadratic and bilinear functionals of $\bmu$ and $\bSigma$:
\[
\mathbf{1}_p^\top \bSigma^{-1} \mathbf{1}_p, \quad
\mathbf{1}_p^\top \bSigma^{-1} \bmu, \quad
\bmu^\top \bSigma^{-1} \bmu.
\]
Our estimation framework can thus be naturally extended to estimate the entire MVF.

To stabilize the estimation of $\mathbf{1}_p^\top \bSigma^{-1} \bmu$ when $\|\mathbf{1}_p\|$ and $\|\bmu\|$ may differ in scale, we normalize these vectors and work with their unit-length versions:
\[
\mathbf{1}_0 = \frac{\mathbf{1}_p}{\sqrt{p}}, 
\qquad
\bmu_0 = \frac{\bmu}{\|\bmu\|}.
\]
Then,
\(\mathbf{1}_p^\top \bSigma^{-1} \bmu 
   = \sqrt{p}\|\bmu\|\,
     \mathbf{1}_0^\top \bSigma^{-1} \bmu_0\),
and $\mathbf{1}_0^\top \bSigma^{-1} \bmu_0$ can be expressed via the polarization identity,
\[
\mathbf{1}_0^\top \bSigma^{-1} \bmu_0
= \tfrac12\Big[
  (\mathbf{1}_0+\bmu_0)^\top \bSigma^{-1} (\mathbf{1}_0+\bmu_0)
  - \mathbf{1}_0^\top \bSigma^{-1}\mathbf{1}_0
  - \bmu_0^\top \bSigma^{-1}\bmu_0
\Big].
\]
Thus, by applying the same estimation procedure to these normalized forms, we can consistently estimate all functionals required to recover the efficient frontier in high-dimensional settings,  
without imposing additional structural assumptions.
\end{remark}

\subsection{Estimating the multiple correlation coefficient}
The multiple correlation coefficient (MCC) quantifies the linear dependence between a univariate response $y$ and a set of predictors $\bx \in \mathbb{R}^p$, defined as
\[
\rho_p \triangleq \max_{\alpha \in \mathbb{R}^p} {\rm Cor}(y,\alpha^\top\bx).
\]
It admits an equivalent representation in terms of variances and covariances:
\begin{align}\label{rhop2}
\rho_p^2
= \frac{\bsigma_{\bx y}^\top \bSigma_{\bx\bx}^{-1} \bsigma_{\bx y}}
       {\sigma_{yy}},
\end{align}
where $\sigma_{yy} = \mathrm{Var}(y)$, $\bSigma_{\bx\bx} = \mathrm{Cov}(\bx)$, and $\bsigma_{\bx y} = \mathrm{Cov}(\bx,y)$.  
In linear regression, $\rho_p^2$ measures the proportion of variance in $y$ explained by $\bx$, thus serving as a key indicator of goodness of fit. 
Accurate estimation of $\rho_p^2$ is therefore essential for statistical inference and predictive modeling.

Suppose we observe a sample of size $n$, consisting of i.i.d.\ pairs 
$\{(y_i,\bx_i): i=1,\ldots,n\}$.
Denote $\y = (y_1,\ldots,y_n)^\top\in\mathbb R^p$ and $\X = (\bx_1,\ldots,\bx_n)\in \mathbb R^{p\times n}$.  
A conventional estimator of $\rho_p^2$ is its empirical analogue, namely the coefficient of determination $R^2$,
\begin{align}\label{eq:R2}
R^2
= \frac{\bs_{\bx y}^\top \bS_{\bx\bx}^{-1} \bs_{\bx y}}{s_{yy}},
\end{align}
where the sample covariance quantities are defined as
\[
s_{yy} = \frac1{n-1} \y^\top \bP_1 \y,\quad
\bS_{\bx\bx} = \frac1{n-1} \X \bP_1 \X^\top,\quad
\bs_{\bx y} = \frac1{n-1} \X \bP_1 \y,
\]
with $\bP_1 = \I_n - \1_n\1_n^\top/n$ denoting the centering projection matrix.  
Here, $s_{yy}$, $\bS_{\bx\bx}$, and $\bs_{\bx y}$ are the unbiased moment estimators of $\sigma_{yy}$, $\bSigma_{\bx\bx}$, and $\bsigma_{\bx y}$, respectively.

The $R^2$ statistic is severely biased in high-dimensional settings where $p$ is comparable to $n$.  
When $p < n$, several bias-corrected estimators have been proposed; see, for example, \citet{Zheng14,Li23,Hong2025}.  
In contrast, for $p > n$, the $R^2$ statistic is no longer well-defined.  
Replacing $\bS_{\bx\bx}^{-1}$ in \eqref{eq:R2} with its Moore–Penrose pseudoinverse $\bS_{\bx\bx}^{+}$ yields a degenerate statistic:
\[
R^2=\frac{\bs_{\bx y}^\top \bS_{\bx\bx}^{+} \bs_{\bx y}}{s_{yy}}
   = \frac{\y^\top \bP_1 \X (\X \bP_1 \X^\top)^{+} \X^\top \bP_1 \y}
          {\y^\top \bP_1 \y}
   \equiv 1.
\]
Indeed, when $p>n$, the fitted response  
\(\hat\y = \bP_1 \X^\top (\X \bP_1 \X^\top)^{+} \X \bP_1 \y\)  
coincides with the projected data \(\bP_1 \y\), leading to a perfect in-sample fit.  
This degeneracy highlights the need for alternative approaches when $p > n$. 
To address this issue, \citet{Kong18} proposed an estimator of $\rho_p^2$ for a linear model based on polynomial approximation.

We next apply the proposed estimation framework to the MCC $\rho_p^2$.  
As shown in \eqref{rhop2}, $\rho_p^2$ can be expressed as a quadratic functional with  
$\ba = \sigma_{yy}^{-1/2}\bsigma_{\bx y}$,  
which enables us to leverage the general results from Section~\ref{results} while incorporating problem-specific adjustments.

Following the same strategy as in Section~\ref{sec:3-1}, we decompose the cross-covariance vector into its scale and direction, which yields the normalized representation
\begin{align*}
\rho_p^2
= \kappa_{\sigma}\,\bsigma_{0}^\top \bSigma_{\bx\bx}^{-1} \bsigma_{0},
\quad \text{where} \quad 
\kappa_{\sigma}=\frac{\bsigma_{\bx y}^\top\bsigma_{\bx y}}{\sigma_{yy}},
\quad 
\bsigma_0=\frac{\bsigma_{\bx y}}{\|\bsigma_{\bx y}\|}.
\end{align*}
The scalar $\kappa_{\sigma}$ and the vector $\bsigma_0$ can be estimated by
\begin{align*}
\hat\kappa_{\sigma}
&=\left|\frac{1}{n(n-1)}
\sum_{i\neq j}
\frac{(y_i-\bar y)(y_j-\bar y) 
(\bx_i-\bar\bx)^\top(\bx_j-\bar\bx)}
      {s_{yy}}\right|,
\quad
\hat\bsigma_0=\frac{\bs_{\bx y}}{\sqrt{s_{yy}\hat\kappa_{\sigma}}},
\end{align*}
where $\bar y=(1/n)\sum_{i=1}^ny_i$ and $\bar\bx=(1/n)\sum_{i=1}^n\bx_i$.
To recover $\bsigma_{0}^\top \bSigma_{\bx\bx}^{-1} \bsigma_{0}$, we analyze the limit of $\hat s_n(z)$. 
An explicit calculation shows that
\begin{align*}
\hat s_n(z) 
&= \hat\bsigma_0^\top \!\left(\bS_n - z\I_p\right)^{-1} \hat\bsigma_0 \\
&= z^2 \um_n^2(z) 
\bsigma_0^\top \!\left(\bS_n - z\I_p\right)^{-1} \bsigma_0
   +\kappa_{\sigma}^{-1}\!\left(1 + z \um_n(z)\right)
   + o_p(1).
\end{align*}
Based on this relationship, we define the bias-corrected function
\begin{align*}
s_{n,\mathrm{MCC}}(z)
=\frac{
\hat\bsigma_{0}^\top\left(\bS_{\bx\bx}-z\I_p\right)^{-1}\hat\bsigma_{0}
-\hat\kappa_{\sigma}^{-1}\!\left(1+z\um_n(z)\right)}
{z^2\um_n^2(z)},
\end{align*}
and use it to construct the moment estimators:
\begin{align}\label{intMCC}
\hat\alpha_{j,\mathrm{MCC}}
= (-1)^j \frac{1}{2\pi\mathrm i}
   \oint_{\mathcal C} 
   \frac{z s_{n,\mathrm{MCC}}(z) \um_n'(z)}{\um_n^{j+1}(z)} dz,
   \qquad j\geq 1.
\end{align}
Combining this with Algorithm~\ref{alg} yields an estimator $\hat{F}^{\bSigma,\bsigma_0}$ of the VESD $F^{\bSigma,\bsigma_0}$.
Finally, we estimate $\rho_p^2$ as
\[
\hat\rho_p^2
  = \hat\kappa_{\sigma}
    \int x^{-1}   d\hat{F}^{\bSigma,\bsigma_0}(x).
\]

\begin{theorem}\label{th-MCC}
Suppose the conditions of Theorem \ref{th-W1distance} are satisfied for the vector $\left(y,\bx^\top\right)^\top$. Then, we have
\[
\hat\rho_p^2 - \rho_p^2 \xrightarrow{i.p.} 0,
\]
as $n,p\to\infty$.
\end{theorem}

\begin{remark}
Theorem~\ref{th-MCC} establishes the consistency of the proposed estimator under high-dimensional asymptotics with $p/n \to c \in (0,\infty)$.  
This ensures valid inference on the MCC in modern high-dimensional settings and addresses the failure of classical methods, such as the standard $R^2$ statistic, which degenerates to $1$ when $p>n$ due to perfect in-sample fitting.
To facilitate implementation, Appendix~\ref{residueMCC} provides Mathematica code for generating the residue-based moment estimators in \eqref{intMCC}.  
\end{remark}

\section{Simulation}
We conduct simulation studies to evaluate the performance of the proposed estimators of the squared optimal Sharpe ratio $\theta_p$ and the squared MCC $\rho_p^2$.  
For comparison, we also consider two shrinkage-based estimators that approximate the population covariance matrix with shrinkage versions of the sample covariance matrix.
The corresponding estimators of $\theta_p$ and $\rho_p^2$ are defined as 
\begin{align*}
\hat\theta_{\mathrm{Sh1}} = \bar\bx^\top \hat\bSigma_{\mathrm{Sh1}}^{-1} \bar\bx,\quad \hat\theta_{\mathrm{Sh2}} = \bar\bx^\top \hat\bSigma_{\mathrm{Sh2}}^{-1} \bar\bx
\end{align*}
and
\begin{align*}
\hat\rho_{\mathrm{Sh1}}^2 = \frac{\bs_{\bx y}^\top \hat\bSigma_{\mathrm{Sh1}}^{-1} \bs_{\bx y}}{s_{yy}},\quad \hat\rho_{\mathrm{Sh2}}^2 = \frac{\bs_{\bx y}^\top \hat\bSigma_{\mathrm{Sh2}}^{-1} \bs_{\bx y}}{s_{yy}},
\end{align*}
where $\hat\bSigma_{\mathrm{Sh1}}$ and $\hat\bSigma_{\mathrm{Sh2}}$ denote the shrinkage estimators of $\bSigma$ in \cite{Ledoit2015} and \cite{Ledoit2018}, respectively.
For the estimation of $\rho_p^2$, we further include the estimator proposed in \citet{Kong18}, denoted by $\hat\rho_{\mathrm{Kong}}^2$.

In implementing the linear program, we set the moment order to $k=4$ and the step size to $h=1/p$. Additionally, we make the following two adjustments:
\begin{itemize}
\item[]{\bf (i) Moment estimation and weighting.} To improve the numerical stability of higher-order moment estimation, we use truncated versions of the moment estimators $\hat\alpha_{j,\mathrm{SR}}$ and $\hat\alpha_{j,\mathrm{MCC}}$ with the truncation parameter $\delta = 0.01$,
and solve the corresponding weighted–objective LP as described in Section \ref{VESD-est}.
\item[]{\bf (ii) Data-driven choice of the support interval.} Noting that the VESDs $F^{\bSigma,\bmu_0}$ and $F^{\bSigma,\bsigma_0}$ are supported on the spectrum of $\bSigma$, we estimate the eigenvalues using the method developed by \citet{Ledoit2015}. Let $\hat\lambda_1 \ge \cdots \ge \hat\lambda_p$ denote the ordered estimates. We then set the working interval to $[a_0,b_0]=[0.8\hat\lambda_{p-1},1.2\hat\lambda_2]$ to mitigate sensitivity to extreme values.
\end{itemize}
For the method of \citet{Kong18}, we set the approximation order to 4 and use the same working interval $[a_0,b_0]$ as in our LP implementation.

We next outline the data-generating settings for the Sharpe ratio and MCC experiments and report the associated numerical results.

\subsection{Estimation of the optimal Sharpe ratio}
The data are generated according to
\begin{align*}
\bx = \bmu + \bSigma^{1/2}\z.
\end{align*}
Two distributional settings are considered for the latent vector $\z$:
\begin{itemize}
\item[]{\bf Model 1 (Independence).} $\z$ consists of i.i.d.\ standard normal entries.
\item[]{\bf Model 2 (Log-concavity).} $\z$ follows an elliptical distribution of the form
\begin{align*}
\z = \frac{1}{\sqrt{p+1}} \xi \bu,
\end{align*}
where $\bu\in \mathbb R^p$ is uniformly distributed on the unit sphere $\mathbb{S}^{p-1}$, independent of $\xi \sim \mathrm{Gamma}(p,1)$.
\end{itemize}
For the covariance structure $\bSigma$, we examine two cases:
\begin{itemize}
\item[]{\bf Case 3.} Diagonal matrix $\bSigma = \mathrm{diag}(\sigma_{11},\ldots,\sigma_{pp})$, with the first $p/2$ diagonal entries equal to $3$ and the remaining $p/2$ equal to $1.5$;  
\item[]{\bf Case 4.} Toeplitz matrix $\bSigma = (\sigma_{ij})$, where $\sigma_{ij} = 2 \cdot 0.3^{|i-j|}$.
\end{itemize}
The mean vector $\bmu$ is defined in two scenarios:
\begin{itemize}
\item[]{\bf Dense setting 2:} All components equal $1/\sqrt{p}$; 
\item[]{\bf Sparse setting 2:} The first entry is 0.6, the second is 0.8, and the rest are zero.
\end{itemize}
The dimensional ratios are set to $c_n \in \{1.25, 1.5\}$ with sample sizes $n \in \{400, 800, 1600\}$, and all results are based on 5000 independent replications.

Tables \ref{t3}-\ref{t4} show the empirical biases and variances of the three estimators $\hat\theta_p$, $\hat\theta_{\mathrm{Sh1}}$ and $\hat\theta_{\mathrm{Sh2}}$.  Across all settings, both the bias and variance of $\hat\theta_p$ decrease as $n$ and $p$ increase, confirming its consistency. In contrast, the biases of $\hat\theta_{\mathrm{Sh1}}$ and $\hat\theta_{\mathrm{Sh2}}$ remain non-negligible even in large
samples, demonstrating that these shrinkage-based estimators are inconsistent.

\setlength{\tabcolsep}{12pt}
\renewcommand{\arraystretch}{1.25}
\begin{table}[htbp]
\centering
\captionsetup{font={small}}
\caption{Empirical biases (variances) of the three estimators of $\theta_p$ from 5000 replications under Model 1.}
\label{t3}
\begin{tabular}{cccc} \Xhline{2pt}
\multicolumn{4}{c}{Case 3 with Dense setting 2} \\ \Xhline{2pt}
\multirow{2}{*}{$(c_n,n)$} & \multicolumn{3}{c}{Method}  \\ \cline{2-4}
& $\hat\theta_p$ & $\hat\theta_{\mathrm{Sh1}}$ & $\hat\theta_{\mathrm{Sh2}}$ \\ \hline
(1.25,400)              & 0.0334(0.0422) & 1.1953(0.0111) & 1.4094(0.0144) \\
(1.25,800)              & 0.0155(0.0219) & 1.1953(0.0056) & 1.4034(0.0073) \\
(1.25,1600)             & 0.0130(0.0123) & 1.1978(0.0029) & 1.4031(0.0037) \\
(1.5,400)               & 0.0331(0.0472) & 1.4468(0.0132) & 1.6941(0.0736) \\
(1.5,800)               & 0.0182(0.0274) & 1.4458(0.0066) & 1.6868(0.0150) \\
(1.5,1600)              & 0.0150(0.0141) & 1.4489(0.0032) & 1.6856(0.0041) \\ \Xhline{2pt}
\multicolumn{4}{c}{Case 3 with Sparse setting 2} \\ \Xhline{2pt}
\multirow{2}{*}{$(c_n,n)$} & \multicolumn{3}{c}{Method}  \\ \cline{2-4}
& $\hat\theta_p$ & $\hat\theta_{\mathrm{Sh1}}$ & $\hat\theta_{\mathrm{Sh2}}$ \\ \hline
(1.25,400)              & 0.0533(0.0250) & 1.3513(0.0128) & 1.5620(0.0164) \\
(1.25,800)              & 0.0341(0.0126) & 1.3513(0.0065) & 1.5564(0.0083) \\
(1.25,1600)             & 0.0275(0.0064) & 1.3525(0.0033) & 1.5548(0.0042) \\
(1.5,400)               & 0.0661(0.0304) & 1.6072(0.0140) & 1.8521(0.0701) \\
(1.5,800)               & 0.0413(0.0155) & 1.6049(0.0072) & 1.8436(0.0166) \\
(1.5,1600)              & 0.0328(0.0080) & 1.6066(0.0035) & 1.8410(0.0045) \\ \Xhline{2pt}
\multicolumn{4}{c}{Case 4 with Dense setting 2} \\ \Xhline{2pt}
\multirow{2}{*}{$(c_n,n)$} & \multicolumn{3}{c}{Method}  \\ \cline{2-4}
& $\hat\theta_p$ & $\hat\theta_{\mathrm{Sh1}}$ & $\hat\theta_{\mathrm{Sh2}}$ \\ \hline
(1.25,400)              & 0.0346(0.0149) & 1.4395(0.0154) & 1.7363(0.0216) \\
(1.25,800)              & 0.0221(0.0076) & 1.4394(0.0076) & 1.7320(0.0105) \\
(1.25,1600)             & 0.0176(0.0037) & 1.4429(0.0039) & 1.7346(0.0053) \\
(1.5,400)               & 0.0459(0.0193) & 1.6972(0.0178) & 2.0476(0.0252) \\
(1.5,800)               & 0.0289(0.0097) & 1.6965(0.0089) & 2.0417(0.0125) \\
(1.5,1600)              & 0.0213(0.0047) & 1.6998(0.0043) & 2.0434(0.0060) \\ \Xhline{2pt}
\multicolumn{4}{c}{Case 4 with Sparse setting 2} \\ \Xhline{2pt}
\multirow{2}{*}{$(c_n,n)$} & \multicolumn{3}{c}{Method}  \\ \cline{2-4}
& $\hat\theta_p$ & $\hat\theta_{\mathrm{Sh1}}$ & $\hat\theta_{\mathrm{Sh2}}$ \\ \hline
(1.25,400)              & 0.0456(0.0312) & 1.3188(0.0139) & 1.6218(0.0196) \\
(1.25,800)              & 0.0209(0.0151) & 1.3184(0.0070) & 1.6169(0.0098) \\
(1.25,1600)             & 0.0100(0.0079) & 1.3201(0.0035) & 1.6174(0.0049) \\
(1.5,400)               & 0.0643(0.0404) & 1.5755(0.0154) & 1.9319(0.0218) \\
(1.5,800)               & 0.0324(0.0194) & 1.5723(0.0078) & 1.9229(0.0110) \\
(1.5,1600)              & 0.0200(0.0102) & 1.5740(0.0039) & 1.9227(0.0054) \\ \Xhline{2pt}
\end{tabular}
\end{table}

\begin{table}[htbp]
\centering
\captionsetup{font={small}}
\caption{Empirical biases (variances) of the three estimators of $\theta_p$ from 5000 replications under Model 2.}
\label{t4}
\begin{threeparttable}
\begin{tabular}{cccc} \Xhline{2pt}
\multicolumn{4}{c}{Case 3 with Dense setting 2} \\ \Xhline{2pt}
\multirow{2}{*}{$(c_n,n)$} & \multicolumn{3}{c}{Method}  \\ \cline{2-4}
& $\hat\theta_p$ & $\hat\theta_{\mathrm{Sh1}}$ & $\hat\theta_{\mathrm{Sh2}}$ \\ \hline
(1.25,400)              & 0.0372(0.0427) & 1.1955(0.0111) & 1.4175(0.0145)  \\
(1.25,800)              & 0.0167(0.0221) & 1.1955(0.0056) & 1.4069(0.0073)  \\
(1.25,1600)             & 0.0125(0.0122) & 1.1979(0.0029) & 1.4040(0.0037)  \\
(1.5,400)               & 0.0354(0.0476) & 1.4469(0.0132) & 1.8217(73.1198*) \\
(1.5,800)               & 0.0207(0.0276) & 1.4460(0.0066) & 1.6941(0.0959)  \\
(1.5,1600)              & 0.0163(0.0144) & 1.4489(0.0032) & 1.6881(0.0041) \\ \Xhline{2pt}
\multicolumn{4}{c}{Case 3 with Sparse setting 2} \\ \Xhline{2pt}
\multirow{2}{*}{$(c_n,n)$} & \multicolumn{3}{c}{Method}  \\ \cline{2-4}
& $\hat\theta_p$ & $\hat\theta_{\mathrm{Sh1}}$ & $\hat\theta_{\mathrm{Sh2}}$ \\ \hline
(1.25,400)              & 0.0546(0.0252) & 1.3512(0.0128) & 1.5696(0.0165)  \\
(1.25,800)              & 0.0353(0.0127) & 1.3514(0.0065) & 1.5598(0.0083)  \\
(1.25,1600)             & 0.0275(0.0063) & 1.3525(0.0033) & 1.5557(0.0041)  \\
(1.5,400)               & 0.0667(0.0308) & 1.6071(0.0140) & 1.9734(65.8021*) \\
(1.5,800)               & 0.0425(0.0156) & 1.6049(0.0072) & 1.8504(0.0854)  \\
(1.5,1600)              & 0.0339(0.0081) & 1.6066(0.0035) & 1.8434(0.0045) \\ \Xhline{2pt}
\multicolumn{4}{c}{Case 4 with Dense setting 2} \\ \Xhline{2pt}
\multirow{2}{*}{$(c_n,n)$} & \multicolumn{3}{c}{Method}  \\ \cline{2-4}
& $\hat\theta_p$ & $\hat\theta_{\mathrm{Sh1}}$ & $\hat\theta_{\mathrm{Sh2}}$ \\ \hline
(1.25,400)              & 0.0365(0.0150) & 1.4390(0.0155) & 1.7427(0.0218) \\
(1.25,800)              & 0.0226(0.0077) & 1.4392(0.0076) & 1.7353(0.0106) \\
(1.25,1600)             & 0.0178(0.0037) & 1.4428(0.0039) & 1.7365(0.0054) \\
(1.5,400)               & 0.0472(0.0196) & 1.6968(0.0178) & 2.0553(0.0255) \\
(1.5,800)               & 0.0289(0.0098) & 1.6964(0.0089) & 2.0458(0.0125) \\
(1.5,1600)              & 0.0216(0.0047) & 1.6998(0.0043) & 2.0453(0.0060) \\ \Xhline{2pt}
\multicolumn{4}{c}{Case 4 with Sparse setting 2} \\ \Xhline{2pt}
\multirow{2}{*}{$(c_n,n)$} & \multicolumn{3}{c}{Method}  \\ \cline{2-4}
& $\hat\theta_p$ & $\hat\theta_{\mathrm{Sh1}}$ & $\hat\theta_{\mathrm{Sh2}}$ \\ \hline
(1.25,400)              & 0.0476(0.0322) & 1.3189(0.0139) & 1.6291(0.0198) \\
(1.25,800)              & 0.0214(0.0150) & 1.3185(0.0070) & 1.6207(0.0098) \\
(1.25,1600)             & 0.0105(0.0079) & 1.3202(0.0035) & 1.6195(0.0049) \\
(1.5,400)               & 0.0660(0.0406) & 1.5756(0.0154) & 1.9405(0.0220) \\
(1.5,800)               & 0.0338(0.0200) & 1.5724(0.0078) & 1.9273(0.0111) \\
(1.5,1600)              & 0.0204(0.0102) & 1.5740(0.0039) & 1.9246(0.0054) \\ \Xhline{2pt}
\end{tabular}
\begin{tablenotes}
\item[*] The large variance is due to rare replications where the smallest eigenvalue estimate $\hat\lambda_p$ from \citet{Ledoit2015} is close to zero, which drives some eigenvalues of $\hat\bSigma_{\mathrm{Sh2}}$ toward zero; see (6.4) and (6.6) in \citet{Ledoit2018}.  
\end{tablenotes}
\end{threeparttable}
\end{table}

\subsection{Estimation of the MCC}
The data are generated from the model
\[
\begin{pmatrix}
y\\
\bx
\end{pmatrix}
= \bSigma_{y\bx}^{1/2}\z_{y\bx},
\]
where $\bSigma_{y\bx}$ is the covariance matrix of $(y,\bx^\top)^\top$ and $\z_{y\bx}$ is a $(p+1)$-dimensional latent vector.  
The distributional settings of $\z_{y\bx}$ follow {\bf Model 1} and {\bf Model 2}, with the dimension $p$ replaced by $p+1$.  
The covariance block $\bSigma_{\bx\bx}$ adopts the same structures as in {\bf Case 3} and {\bf Case 4}, while the direction vector $\bsigma_{\bx y}$ is defined by {\bf Dense setting 2} and {\bf Sparse setting 2}.  
We fix $\sigma_{yy}=1$ throughout, and the dimensional settings are identical to those in the Sharpe ratio experiment.

Tables \ref{t5}--\ref{t6} present the empirical biases and variances of the four estimators $\hat\rho_p^2$, $\hat\rho_{\mathrm{Sh1}}^2$, $\hat\rho_{\mathrm{Sh2}}^2$ and $\hat\rho_{\mathrm{Kong}}^2$, based on 5000 replications.  
The results show that both $\hat\rho_p^2$ and $\hat\rho_{\mathrm{Kong}}^2$ are consistent, as their biases and variances decrease with increasing $(n,p)$, while the shrinkage-based estimators remain biased. Although $\hat\rho_{\mathrm{Kong}}^2$ exhibits small bias, its variance is considerably larger than that of the other competitors. As a result, the proposed estimator $\hat\rho_p^2$ attains the smallest mean squared error among all methods.

\begin{table}[htbp]
\centering
\captionsetup{font={small}}
\caption{Empirical biases (variances) of the four estimators of $\rho_p^2$ from 5000 replications under Model 1.}
\label{t5}
	\scalebox{0.87}{
\begin{tabular}{ccccc} \Xhline{2pt}
\multicolumn{5}{c}{Case 3 with Dense setting 2} \\ \Xhline{2pt}
\multirow{2}{*}{$(c_n,n)$} & \multicolumn{4}{c}{Method}  \\ \cline{2-5}
& $\hat\rho_p^2$ & $\hat\rho_{\mathrm{Sh1}}^2$ & $\hat\rho_{\mathrm{Sh2}}^2$ & $\hat\rho_{\mathrm{Kong}}^2$ \\ \hline
(1.25,400)              & 0.0338(0.0246) & 0.9997(0.0067) & 1.1539(0.0077) & 0.0020(0.2228) \\
(1.25,800)              & 0.0244(0.0133) & 1.0012(0.0033) & 1.1539(0.0039) & 0.0133(0.1208) \\
(1.25,1600)             & 0.0158(0.0077) & 1.0023(0.0018) & 1.1542(0.0020) & 0.0053(0.0631) \\
(1.5,400)               & 0.0314(0.0281) & 1.2243(0.0074) & 1.4041(0.0085) & 0.0029(0.4282) \\
(1.5,800)               & 0.0216(0.0170) & 1.2254(0.0038) & 1.4033(0.0043) & -0.00004(0.2317) \\
(1.5,1600)              & 0.0175(0.0098) & 1.2268(0.0019) & 1.4039(0.0022) & 0.0062(0.1191) \\ \Xhline{2pt}
\multicolumn{5}{c}{Case 3 with Sparse setting 2} \\ \Xhline{2pt}
\multirow{2}{*}{$(c_n,n)$} & \multicolumn{4}{c}{Method}  \\ \cline{2-5}
& $\hat\rho_p^2$ & $\hat\rho_{\mathrm{Sh1}}^2$ & $\hat\rho_{\mathrm{Sh2}}^2$ & $\hat\rho_{\mathrm{Kong}}^2$ \\ \hline
(1.25,400)              & 0.0497(0.0130) & 1.1589(0.0075) & 1.3106(0.0087) & -0.0052(0.2471) \\
(1.25,800)              & 0.0342(0.0067) & 1.1615(0.0037) & 1.3117(0.0043) & 0.0040(0.1331) \\
(1.25,1600)             & 0.0264(0.0034) & 1.1622(0.0019) & 1.3115(0.0022) & -0.0073(0.0688) \\
(1.5,400)               & 0.0571(0.0169) & 1.3837(0.0087) & 1.5614(0.0101) & -0.0032(0.4486) \\
(1.5,800)               & 0.0385(0.0088) & 1.3862(0.0042) & 1.5618(0.0048) & -0.0124(0.2472) \\
(1.5,1600)              & 0.0284(0.0043) & 1.3871(0.0021) & 1.5619(0.0024) & -0.0047(0.1281) \\ \Xhline{2pt}
\multicolumn{5}{c}{Case 4 with Dense setting 2} \\ \Xhline{2pt}
\multirow{2}{*}{$(c_n,n)$} & \multicolumn{4}{c}{Method}  \\ \cline{2-5}
& $\hat\rho_p^2$ & $\hat\rho_{\mathrm{Sh1}}^2$ & $\hat\rho_{\mathrm{Sh2}}^2$ & $\hat\rho_{\mathrm{Kong}}^2$ \\ \hline
(1.25,400)              & 0.0292(0.0069) & 1.1357(0.0069) & 1.3754(0.0091) & -0.0050(0.0822) \\
(1.25,800)              & 0.0184(0.0033) & 1.1369(0.0034) & 1.3754(0.0045) & -0.0068(0.0392) \\
(1.25,1600)             & 0.0138(0.0018) & 1.1383(0.0018) & 1.3761(0.0023) & -0.0089(0.0209) \\
(1.5,400)               & 0.0389(0.0090) & 1.3471(0.0075) & 1.6303(0.0099) & -0.0087(0.1364) \\
(1.5,800)               & 0.0235(0.0043) & 1.3476(0.0037) & 1.6290(0.0049) & -0.0131(0.0670) \\
(1.5,1600)              & 0.0184(0.0023) & 1.3496(0.0019) & 1.6303(0.0025) & -0.0103(0.0333) \\ \Xhline{2pt}
\multicolumn{5}{c}{Case 4 with Sparse setting 2} \\ \Xhline{2pt}
\multirow{2}{*}{$(c_n,n)$} & \multicolumn{4}{c}{Method}  \\ \cline{2-5}
& $\hat\rho_p^2$ & $\hat\rho_{\mathrm{Sh1}}^2$ & $\hat\rho_{\mathrm{Sh2}}^2$ & $\hat\rho_{\mathrm{Kong}}^2$ \\ \hline
(1.25,400)              & 0.0468(0.0165) & 0.9999(0.0056) & 1.2452(0.0074) & 0.0030(0.0732) \\
(1.25,800)              & 0.0266(0.0081) & 1.0019(0.0028) & 1.2466(0.0037) & 0.0105(0.0348) \\
(1.25,1600)             & 0.0109(0.0038) & 1.0024(0.0014) & 1.2465(0.0019) & 0.0079(0.0188) \\
(1.5,400)               & 0.0590(0.0219) & 1.2082(0.0065) & 1.4962(0.0087) & 0.0059(0.1216) \\
(1.5,800)               & 0.0363(0.0116) & 1.2092(0.0031) & 1.4959(0.0041) & 0.0033(0.0597) \\
(1.5,1600)              & 0.0201(0.0053) & 1.2103(0.0016) & 1.4963(0.0021) & 0.0045(0.0319) \\ \Xhline{2pt}
\end{tabular}}
\end{table}

\begin{table}[htbp]
\centering
\captionsetup{font={small}}
\caption{Empirical biases (variances) of the four estimators of $\rho_p^2$ from 5000 replications under Model 2.}
\label{t6}
	\scalebox{0.87}{
\begin{tabular}{ccccc} \Xhline{2pt}
\multicolumn{5}{c}{Case 3 with Dense setting 2} \\ \Xhline{2pt}
\multirow{2}{*}{$(c_n,n)$} & \multicolumn{4}{c}{Method}  \\ \cline{2-5}
& $\hat\rho_p^2$ & $\hat\rho_{\mathrm{Sh1}}^2$ & $\hat\rho_{\mathrm{Sh2}}^2$ & $\hat\rho_{\mathrm{Kong}}^2$ \\ \hline
(1.25,400)              & 0.0259(0.0241) & 0.9967(0.0066) & 1.1568(0.0077) & 0.0011(0.2065) \\
(1.25,800)              & 0.0194(0.0132) & 0.9998(0.0033) & 1.1555(0.0039) & 0.0130(0.1147) \\
(1.25,1600)             & 0.0130(0.0078) & 1.0015(0.0017) & 1.1549(0.0020) & 0.0046(0.0630) \\
(1.5,400)               & 0.0235(0.0275) & 1.2205(0.0074) & 1.4073(0.0085) & 0.0050(0.3948) \\
(1.5,800)               & 0.0181(0.0168) & 1.2232(0.0038) & 1.4047(0.0043) & 0.0010(0.2221) \\
(1.5,1600)              & 0.0160(0.0098) & 1.2256(0.0019) & 1.4045(0.0022) & 0.0063(0.1181) \\ \Xhline{2pt}
\multicolumn{5}{c}{Case 3 with Sparse setting 2} \\ \Xhline{2pt}
\multirow{2}{*}{$(c_n,n)$} & \multicolumn{4}{c}{Method}  \\ \cline{2-5}
& $\hat\rho_p^2$ & $\hat\rho_{\mathrm{Sh1}}^2$ & $\hat\rho_{\mathrm{Sh2}}^2$ & $\hat\rho_{\mathrm{Kong}}^2$ \\ \hline
(1.25,400)              & 0.0433(0.0126) & 1.1554(0.0075) & 1.3130(0.0087) & -0.0055(0.2283) \\
(1.25,800)              & 0.0316(0.0066) & 1.1598(0.0037) & 1.3129(0.0043) & 0.0034(0.1267) \\
(1.25,1600)             & 0.0245(0.0033) & 1.1613(0.0019) & 1.3120(0.0022) & -0.0073(0.0683) \\
(1.5,400)               & 0.0520(0.0164) & 1.3796(0.0087) & 1.5641(0.0102) & -0.0012(0.4088) \\
(1.5,800)               & 0.0355(0.0086) & 1.3839(0.0041) & 1.5630(0.0048) & -0.0133(0.2349) \\
(1.5,1600)              & 0.0270(0.0043) & 1.3860(0.0021) & 1.5625(0.0025) & -0.0048(0.1277) \\ \Xhline{2pt}
\multicolumn{5}{c}{Case 4 with Dense setting 2} \\ \Xhline{2pt}
\multirow{2}{*}{$(c_n,n)$} & \multicolumn{4}{c}{Method}  \\ \cline{2-5}
& $\hat\rho_p^2$ & $\hat\rho_{\mathrm{Sh1}}^2$ & $\hat\rho_{\mathrm{Sh2}}^2$ & $\hat\rho_{\mathrm{Kong}}^2$ \\ \hline
(1.25,400)              & 0.0251(0.0067) & 1.1332(0.0068) & 1.3777(0.0090) & -0.0070(0.0787) \\
(1.25,800)              & 0.0166(0.0033) & 1.1357(0.0034) & 1.3766(0.0045) & -0.0067(0.0380) \\
(1.25,1600)             & 0.0128(0.0017) & 1.1376(0.0018) & 1.3767(0.0023) & -0.0094(0.0207) \\
(1.5,400)               & 0.0346(0.0087) & 1.3440(0.0074) & 1.6328(0.0098) & -0.0069(0.1306) \\
(1.5,800)               & 0.0213(0.0042) & 1.3459(0.0037) & 1.6301(0.0049) & -0.0130(0.0656) \\
(1.5,1600)              & 0.0173(0.0022) & 1.3487(0.0019) & 1.6309(0.0026) & -0.0107(0.0326) \\ \Xhline{2pt}
\multicolumn{5}{c}{Case 4 with Sparse setting 2} \\ \Xhline{2pt}
\multirow{2}{*}{$(c_n,n)$} & \multicolumn{4}{c}{Method}  \\ \cline{2-5}
& $\hat\rho_p^2$ & $\hat\rho_{\mathrm{Sh1}}^2$ & $\hat\rho_{\mathrm{Sh2}}^2$ & $\hat\rho_{\mathrm{Kong}}^2$ \\ \hline
(1.25,400)              & 0.0379(0.0161) & 0.9975(0.0056) & 1.2477(0.0074) & 0.0016(0.0704) \\
(1.25,800)              & 0.0219(0.0079) & 1.0007(0.0028) & 1.2478(0.0037) & 0.0105(0.0336) \\
(1.25,1600)             & 0.0084(0.0038) & 1.0018(0.0014) & 1.2470(0.0019) & 0.0078(0.0185) \\
(1.5,400)               & 0.0509(0.0217) & 1.2050(0.0065) & 1.4987(0.0087) & 0.0083(0.1169) \\
(1.5,800)               & 0.0328(0.0115) & 1.2075(0.0031) & 1.4971(0.0041) & 0.0037(0.0584) \\
(1.5,1600)              & 0.0177(0.0052) & 1.2095(0.0016) & 1.4970(0.0021) & 0.0045(0.0314) \\ \Xhline{2pt}
\end{tabular}}
\end{table}

\appendix
\section{Calculation for the contour integrals}\label{residue}
\subsection{Calculation for \eqref{hatalpha}}\label{sec-hatalpha}
Let $\psi=\min\left\{p,n-1\right\}$, and denote by $\lambda_1^{\bS_n}>\lambda_2^{\bS_n}>\dots>\lambda_\psi^{\bS_n}$ the nonzero eigenvalues of $\bS_n$. Let $\eta_1>\dots>\eta_\psi$ be the zeros of $\um_n(x)$,
which satisfy the interlacing inequalities:
$$
\lambda_1^{\bS_n}>\eta_1>\lambda_2^{\bS_n}>\eta_2>\dots>\lambda_\psi^{\bS_n}>\eta_\psi.
$$
Define $f_{jn}(z)=zs_n(z)\um_n'(z)/\um_n^{j+1}(z)$. It is straightforward to verify that all poles of $f_{jn}$ lie in the set $\{\lambda_1^{\bS_n},\dots,\lambda_\psi^{\bS_n},\eta_1,\dots,\eta_\psi\}$. Hence, by the residue theorem, the contour integral in \eqref{hatalpha} can be expressed as
\begin{align*}
\frac{1}{2\pi \mathrm i}\oint_{\mathcal C}\frac{zs_{n}(z)\um_n'(z)}{\um_n^{j+1}(z)}dz
=\sum_{i=1}^\psi\text{Res}(f_{jn},\lambda_i^{\bS_n})+\sum_{i=1}^\psi\text{Res}(f_{jn},\eta_i).
\end{align*}
All residues admit closed-form expressions. 
Those at $\{\lambda_i^{\bS_n}\}$ are relatively simple. In particular,
\begin{align*}
\text{Res}(f_{jn},\lambda_i)=-n\lambda_i^{\bS_n}(\ba^\top\bv_i)^2\mathbb{I}(j=1),
\end{align*}
where $\bv_i$ is the eigenvector of $\bS_n$ associated with $\lambda_i^{\bS_n}$. 
In contrast, the residues at $\{\eta_i\}$ involve more complicated formulas. 
For brevity, we omit the explicit expressions here and instead provide \texttt{Mathematica} code for their computation.
{\footnotesize
\begin{verbatim}
j = 1; (* the order of moment *)
f = (z - eta) ^ (j + 1) * z * sn[z] * D[mn[z], z] / (mn[z]) ^ (j + 1);
D[f, {z, j}];
D[% * mn[z] ^ (2 j + 1), {z, 2 j + 1}] /. z -> eta;
D[mn[z], z] ^ (2 j + 1) (j)! (2 j + 1)! /. z -> eta;
Simplify[%% / %, mn[eta] == 0]
\end{verbatim}}

\subsection{Calculation for \eqref{intSR}}\label{residueSR}
Applying the change of variables $u_n(z)=-1/\um_n(z)$ and using the Cauchy integral formula, we obtain
\begin{align}\label{residue0}
\frac{1}{2\pi \mathrm i}\oint_{\mathcal C}\frac{\um_n'(z)}{\um_n^{j+2}(z)}dz=(-1)^j\frac{1}{2\pi \mathrm i}\oint_{\tilde{\mathcal C}}u^j_ndu_n=0,
\end{align}
where $\tilde{\mathcal C}$ is the image of $\mathcal C$ and does not enclose any poles of $u_n^j$.
Therefore, the contour integral in \eqref{intSR} can be decomposed as
\begin{align*}
\frac{1}{2\pi \mathrm i}\oint_{\mathcal C}\frac{zs_{n,\mathrm{SR}}(z)\um_n'(z)}{\um_n^{j+1}(z)}dz=&\frac{\hat\kappa_{\bmu}^{-1}}{2\pi \mathrm i}\oint_{\mathcal C}\frac{z\bar s_n(z)\um_n'(z)}{\um_n^{j+1}(z)}dz+\frac{\hat\kappa_{\bmu}^{-1}}{2\pi \mathrm i}\oint_{\mathcal C}\frac{z\um_n'(z)}{\um_n^{j+1}(z)}dz\\
\triangleq&\hat\kappa_{\bmu}^{-1}\left[C_1(z)+C_2(z)\right],
\end{align*}
where $\bar s_n(z)=\bar\bx^\top\left(\bS_n-z\I_p\right)^{-1}\bar\bx$.
The calculation of $C_1(z)$ follows exactly the same steps as in Appendix~\ref{sec-hatalpha}, except that $\ba$ is replaced by $\bar\bx$.  
For the derivation of $C_2(z)$, we refer to the detailed analysis in \citet{Li14}.

\subsection{Calculation for \eqref{intMCC}}\label{residueMCC}
The contour integral in \eqref{intMCC} can be evaluated analogously to the Sharpe ratio case.  
Using the identity in \eqref{residue0}, we obtain
\begin{align*}
\frac{1}{2\pi \mathrm i}\oint_{\mathcal C}\frac{zs_{n,\mathrm{MCC}}(z)\um_n'(z)}{\um_n^{j+1}(z)}dz
=\frac{\hat\kappa_{\sigma}^{-1}}{2\pi \mathrm i}\oint_{\mathcal C}\frac{\check s_n(z)\um_n'(z)}{z\um_n^{j+3}(z)}dz
=\hat\kappa_{\sigma}^{-1}\sum_{i=1}^\psi\text{Res}(\check f_{jn},\eta_i),
\end{align*}
where 
\begin{align*}
\check s_n(z)=\frac{\bs_{\bx y}^\top\left(\bS_{\bx\bx}-z\I_p\right)^{-1}\bs_{\bx y}}{s_{yy}}-1,\quad\check f_{jn}(z)=\frac{\check s_n(z)\um_n'(z)}{z\um_n^{j+3}(z)}.
\end{align*}
All residues can be computed using the same \texttt{Mathematica} code as in the Sharpe ratio case.
{\footnotesize
	\begin{verbatim}
		j = 1; (* the order of moment *)
		f = (z - eta)^(j + 3) * D[mn[z], z] * sn[z] / z / (mn[z]) ^ (j + 3);
		D[f, {z, j + 2}];
		D[% * mn[z]^(2 j + 5), {z, 2 j + 5}] /. z -> eta;
		D[mn[z], z] ^ (2 j + 5) (j + 2)! (2 j + 5)! /. z -> eta;
		Simplify[%% / %, mn[eta] == 0]
\end{verbatim}}

\begin{funding}
Weiming Li’s research is supported by NSFC (No. 11971293, 12141107). Guangming Pan's research is supported by MOE-T2EP20123-0007.
\end{funding}

\begin{supplement}
\stitle{Supplement to ``High-Dimensional Precision Matrix Quadratic Forms: Estimation Framework for $p > n$"}
\sdescription{This supplementary material provides detailed proofs of all theorems and includes the derivation of \eqref{quad-MPinverse}.}
\end{supplement}

\bibliography{refe.bib}

@book{Anderson03,
  author = {Anderson, T. W.},
  year = {2003},
  title = {An Introduction to Multivariate Statistical Analysis},
  edition={3rd},
  publisher={Wiley},
  address={New York},
}

@book{Bai10,
  author = {Bai, Z. D. and Silverstein, J. W.},
  year = {2010},
  title = {Spectral Analysis of Large Dimensional Random Matrices},
  edition={2nd},
  publisher={Springer},
  address={New York},
}

@article{BaiZhou08,
author = {Bai, Z. D. and Zhou, W.},
year = {2008},
pages = {425-442},
title = {Large sample covariance matrices without independence structures in columns},
volume = {18},
journal = {Statistica Sinica},
}

@article{Kong17,
author = {Kong, W. H. and Valiant, G.},
title = {Spectrum estimation from samples},
volume = {45},
journal = {The Annals of Statistics},
pages = {2218-2247},
year = {2017},
}

@book{SignalProcessing2018, 
address = {Cambridge}, 
title = {Robust Statistics for Signal Processing}, 
publisher = {Cambridge University Press}, 
author = {Zoubir, Abdelhak M. and Koivunen, Visa and Ollila, Esa and Muma, Michael}, 
year = {2018},
}

@book{DiscriminantAnalysis2004,
  author = {Mclachlan, G.},
  year = {2004},
  publisher = {John Wiley \& Sons, Inc.},
  address  = {Hoboken, New Jersey},
  title = {Discriminant Analysis and Statistical Pattern Recognition},
}

@article{MD-SVM2007,
author = {Wang, D. F. and Yeung, D. S. and Tsang, E. C. C.},
year = {2007},
pages = {1453-1462},
title = {Weighted Mahalanobis distance kernels for support vector machines},
volume = {18},
journal = {IEEE Transactions on Neural Networks},
}

@article{BaiPortfolio2009,
author = {Bai, Z. D. and Liu, H. X. and Wong, Wing-Keung},
year = {2009},
pages = {639 - 667},
title = {Enhancement of the applicability of {M}arkowitz's portfolio optimization by itilizing Random Matrix Theory},
volume = {19},
journal = {Mathematical Finance},
}

@article{KanZhou2007, 
title={Optimal Portfolio Choice with Parameter Uncertainty}, 
volume={42},
journal={Journal of Financial and Quantitative Analysis}, 
author={Kan, R. and Zhou, G. F.}, 
year={2007}, 
pages={621–656},
}

@article{pan2014comparison,
  title = {Comparison between two types of large sample covariance matrices},
  author = {Pan, G. M.},
  journal = {Annales de l'Institut Henri Poincaré, Probabilités et Statistiques},
  volume = {50},
  pages = {655--677},
  year = {2014},
}

@article{CLIME2011,
author = {Cai, T. and Liu, W. D. and Luo, X.},
year = {2011},
pages = {594-607},
title = {A Constrained $\ell_1$ Minimization Approach to Sparse Precision Matrix Estimation},
volume = {106},
journal = {Journal of the American Statistical Association},
}

@article{friedman2008sparse,
author = {Friedman, J. and Hastie, T. and Tibshirani, R.},
year = {2008},
pages = {432-441},
title = {Sparse inverse covariance estimation with the graphical LASSO},
volume = {9},
journal = {Biostatistics},
}

@article{scaleLasso2013,
author = {Sun, T. N. and Zhang, C.-H.},
year = {2013},
pages = {3385-3418},
title = {Sparse Matrix Inversion with Scaled Lasso},
volume = {14},
journal = {Journal of Machine Learning Research}
}

@article{CARE2025,
author = {Zhang, S. C. and Wang, H. Y. and Lin, W.},
title = {{CARE}: Large Precision Matrix Estimation for Compositional Data},
journal = {Journal of the American Statistical Association},
volume = {120},
number = {549},
pages = {305-317},
year = {2025},
}

@article{Fan2016,
author = {Fan, Y. Y. and Lv, J. C.},
title = {Innovated scalable efficient estimation in ultra-large Gaussian graphical models},
volume = {44},
journal = {The Annals of Statistics},
pages = {2098-2126},
year = {2016},
}

@article{Ledoit2017,
    author = {Ledoit, O. and Wolf, M.},
    title = {Nonlinear Shrinkage of the Covariance Matrix for Portfolio Selection: Markowitz Meets Goldilocks},
    journal = {The Review of Financial Studies},
    volume = {30},
    pages = {4349-4388},
    year = {2017},
}

@article{Ledoit2012,
author = {Ledoit, O. and Wolf, M.},
title = {Nonlinear shrinkage estimation of large-dimensional covariance matrices},
volume = {40},
journal = {The Annals of Statistics},
pages = {1024-1060},
year = {2012},
}

@article{Ledoit2004,
author = {Ledoit, O. and Wolf, M.},
title = {A well-conditioned estimator for large-dimensional covariance matrices},
journal = {Journal of Multivariate Analysis},
volume = {88},
pages = {365-411},
year = {2004},
}

@article{Fan2008,
title = {High dimensional covariance matrix estimation using a factor model},
journal = {Journal of Econometrics},
volume = {147},
pages = {186-197},
year = {2008},
author = {Fan, J. Q. and Fan, Y. Y. and Lv, J. C.},
}

@article{Fan2013,
    author = {Fan, J. Q. and Liao, Y. and Mincheva, M.},
    title = {Large Covariance Estimation by Thresholding Principal Orthogonal Complements},
    journal = {Journal of the Royal Statistical Society Series B: Statistical Methodology},
    volume = {75},
    pages = {603-680},
    year = {2013},
}

@article{Daniele2025,
    author = {Daniele, M. and Pohlmeier, W. and Zagidullina, A.},
    title = {A Sparse Approximate Factor Model for High-Dimensional Covariance Matrix Estimation and Portfolio Selection},
    journal = {Journal of Financial Econometrics},
    volume = {23},
    pages = {nbae017},
    year = {2025},
}

@article{fan2021optimal,
  title = {Optimal estimation of functionals of high-dimensional mean and covariance matrix},
  author = {Fan, J. Q. and Weng, H. L. and Zhou, Y. F.},
  year = {2021},
  journal={arXiv preprint arXiv:1908.07460}
}

@article{Hong2025,
author = {Hong, S. Z. and Li, W. M. and Liu, Q. and Zhang, Y. C.},
title = {An Adaptive Adjustment to the {$R^2$} Statistic in High-Dimensional Elliptical Models},
journal = {Journal of the American Statistical Association},
volume = {120},
pages = {2372-2384},
year = {2025},
}

@article{Lu2024,
author = {Lu, Y. H. and Yang, Y. R. and Zhang, T.},
year = {2024},
title = {Double Descent in Portfolio Optimization: Dance between Theoretical Sharpe Ratio and Estimation Accuracy},
journal= {arXiv preprint arXiv:2411.18830},
}

@article{Bai2007,
author = {Bai, Z. D. and Miao, B. Q. and Pan, G. M.},
title = {On asymptotics of eigenvectors of large sample covariance matrix},
volume = {35},
journal = {The Annals of Probability},
pages = {1532-1572},
year = {2007},
}

@article{SC95,
title = {Analysis of the Limiting Spectral Distribution of Large Dimensional Random Matrices},
journal = {Journal of Multivariate Analysis},
volume = {54},
pages = {295-309},
year = {1995},
author = {Silverstein, J. W. and Choi, S. I.},
}

@article{portfolio1952,
author = {Markowitz, H.},
title = {Portfolio Selection},
journal = {The Journal of Finance},
volume = {7},
pages = {77-91},
year = {1952}
}

@article{Ao2018,
    author = {Ao, M. M. and Li, Y. Y. and Zheng, X. H.},
    title = {Approaching Mean-Variance Efficiency for Large Portfolios},
    journal = {The Review of Financial Studies},
    volume = {32},
    pages = {2890-2919},
    year = {2018},
}

@article{Li14,
author = {Li, W. M. and Yao, J. F.},
year = {2014},
pages = {919-936},
title = {A local moment estimator of the spectrum of a large dimensional covariance matrix},
volume = {24},
journal = {Statistica Sinica},
}

@article{Li23,
  author={Li, W. M. and Hong, S. Z.},
  journal={Statistica Sinica},
  title={{CLT} for high-dimensional {$R^2$} statistics under a general independent components model},
  year={2024},
  volume = {34},
  pages = {2265-2275},
}

@article{MP67,
  title={The distribution of eigenvalues in certain sets of random matrices},
  author={Mar\v{c}enko, V. A. and Pastur, L. A.},
  journal={Math. USSR-Sbornik},
  volume={1},
  pages={457-483},
  year={1967}
}

@article{S95,
  author = {Silverstein, J. W.},
  title = {Strong convergence of the empirical distribution of eigenvalues of large-dimensional random matrices},
  journal = {Journal of Multivariate Analysis},
  volume = {55},
  year = {1995},
  pages = {331-339},
}

@article{Zheng14,
  title={Inference on multiple correlation coefficients with moderately high dimensional data},
  author={Zheng, S. R. and Jiang, D. D. and Bai, Z. D. and He, X. M.},
  journal={Biometrika},
  volume={101},
  year={2014},
  pages={748-754},
}

@article {Tyler1987,
    AUTHOR = {Tyler, David E.},
     TITLE = {A distribution-free {$M$}-estimator of multivariate scatter},
  JOURNAL = {The Annals of Statistics},
    VOLUME = {15},
      YEAR = {1987},
     PAGES = {234--251},
      ISSN = {0090-5364},
   MRCLASS = {62H12 (62F35 62G05)},
  MRNUMBER = {885734},
MRREVIEWER = {Nariaki Sugiura},
       DOI = {10.1214/aos/1176350263},
       URL = {https://doi.org/10.1214/aos/1176350263},
}

@article{locantore1999robust,
  title={Robust principal component analysis for functional data},
  author={Locantore, N. and Marron, J. and Simpson, D. and Tripoli, N. and Zhang, J. and Cohen, K. and Boente, G. and Fraiman, R. and Brumback, B.and Croux, C. and others},
  journal={Test},
  volume={8},
  pages={1-73},
  year={1999},
}

@article{Ledoit2015,
title = {Spectrum estimation: A unified framework for covariance matrix estimation and PCA in large dimensions},
journal = {Journal of Multivariate Analysis},
volume = {139},
pages = {360-384},
year = {2015},
author = {Ledoit, O. and Wolf, M.},
}

@article{Ledoit2018,
author = {Ledoit, O. and Wolf, M.},
title = {Optimal estimation of a large-dimensional covariance matrix under {S}tein's loss},
volume = {24},
journal = {Bernoulli},
number = {4B},
pages = {3791-3832},
year = {2018},
}

@article{Merton1972, 
title = {An Analytic Derivation of the Efficient Portfolio Frontier}, 
volume = {7},
journal = {Journal of Financial and Quantitative Analysis},
author = {Merton, R.},
year = {1972}, 
pages = {1851–1872},
}

@article{Karoui2010,
author = {El Karoui, N.},
title = {High-dimensionality effects in the {M}arkowitz problem and other quadratic programs with linear constraints: {R}isk underestimation},
volume = {38},
journal = {The Annals of Statistics},
pages = {3487-566},
year = {2010},
}

@article{Fan2018,
author = {Fan, J. Q. and Liu, H. and Wang, W. C.},
title = {Large covariance estimation through elliptical factor models},
volume = {46},
journal = {The Annals of Statistics},
pages = {1383-1414},
year = {2018},
}

@book{Campbell1997, 
address = {Princeton}, 
title = {The Econometrics of Financial Markets},
publisher = {Princeton University Press}, 
author = {Campbell, J. Y. and Lo, A. W. and MacKinlay, A. C.}, 
year = {1997},
}

@article{Meng2025,
author = {Meng, X. R. and Cao, Y. and Wang, W. C.},
title = {Estimation of Out-of-Sample Sharpe Ratio for High Dimensional Portfolio Optimization},
journal = {Journal of the American Statistical Association, To appear},
year = {2025},
}

@article{Kan2024,
title = {In-sample and out-of-sample Sharpe ratios of multi-factor asset pricing models},
journal = {Journal of Financial Economics},
volume = {155},
pages = {103837},
year = {2024},
author = {Kan, R. and Wang, X. L. and Zheng, X. H.},
}

@article{Kong18,
author = {Kong, W. H. and Valiant, G.},
year = {2018},
volume = {31},
pages = {5455–5464},
title = {Estimating Learnability in the Sublinear Data Regime},
journal = {Advances in Neural Information Processing Systems},
}

\end{document}